\documentclass[lettersize,journal]{IEEEtran}
\usepackage[utf8]{inputenc}
\usepackage{textcomp}
\DeclareUnicodeCharacter{2212}{\textminus}
\usepackage[table]{xcolor}

\colorlet{shadecolor}{yellow}
\usepackage[pdftex]{graphicx}

\graphicspath{{Figure/}}
\DeclareGraphicsExtensions{.pdf,.jpeg,.png}

\usepackage{amsmath,amsfonts}
\usepackage{algorithmic}
\usepackage{algorithm}
\usepackage[caption=false,font=normalsize,labelfont=sf,textfont=sf]{subfig}
\usepackage{stfloats}
\usepackage{verbatim}
\usepackage{mdwmath}
\usepackage{mdwtab}
\usepackage{eqparbox}
\usepackage{url}
\usepackage{cite}
\usepackage{multicol}
\usepackage{float}
\usepackage{bbm}
\usepackage{balance}
\usepackage{tabularx}
\usepackage{array}         
\usepackage{multirow}
\newcolumntype{L}{>{\raggedright\arraybackslash}X} 
\newcolumntype{C}{>{\centering\arraybackslash}X}  
\newcolumntype{R}{>{\raggedleft\arraybackslash}X} 
\newcommand{\best}[1]{\textbf{#1}}                 
\newcommand{\catbest}[1]{\underline{#1}}           
\newcommand{\bestcat}[1]{\textbf{\underline{#1}}}  

\usepackage{amssymb}
\usepackage{pifont}

\usepackage{booktabs,multirow,makecell,threeparttable}
\usepackage[table]{xcolor} 
\definecolor{LightGray}{gray}{0.92}

\begin{document}
    \title{SelectTSL: Prompt-Guided Selective Target Sound Localization in Complex Scenarios}
    \author{Ziyang Jiang, \textit{Student Member, IEEE}, Yu Chen, Zexu Pan, \textit{Member, IEEE}, Xinyuan Qian, \textit{Senior Member, IEEE}, Bowen Xing,  Ivor W. Tsang, \textit{Fellow, IEEE}, Xu-Cheng Yin, \textit{Senior Member, IEEE}, Haizhou Li, \textit{Fellow, IEEE}}

\maketitle

\begin{abstract}
Humans can selectively attend to a target sound and estimate its direction in complex scenarios, whereas such selective localization remains challenging for current deep learning based systems. Sound source localization (SSL) has achieved remarkable success with deep learning, yet most methods localize all active sources without selectivity. Conversely, target sound extraction (TSE) extracts sources using multimodal prompts but typically fails to preserve the multichannel spatial information required for accurate localization. To bridge this gap, we formulate the task of prompt-guided selective target sound localization and propose SelectTSL, an end-to-end architecture that localizes only the user-specified target in multi-source acoustic scenes. 
Specifically, we design an \textit{target-aware selective localization strategy} that employs a Prompt-Guided Selective Attention Module (PGSA) to generate prompt-informed embeddings. These embeddings guide an inter-channel phase difference (IPD) enhancer to refine raw phase cues, fusing with target magnitudes to jointly estimate direction of arrival (DoA) and target-source cardinality (i.e., the number of target sound sources).
This coupled design effectively focuses on the user-specified target spatial cues for selective localization and also handles time-varying numbers of target sources. Extensive experiments on
both synthetic data
and real-world recordings demonstrate that our proposed method consistently outperforms other baselines and exhibits robust generalization to real acoustic environments.
\textcolor{magenta}{Dataset and code will be released.}

\end{abstract}

\section{Introduction}

\begin{figure}[!t]
    \centering
    \includegraphics[width=\linewidth]{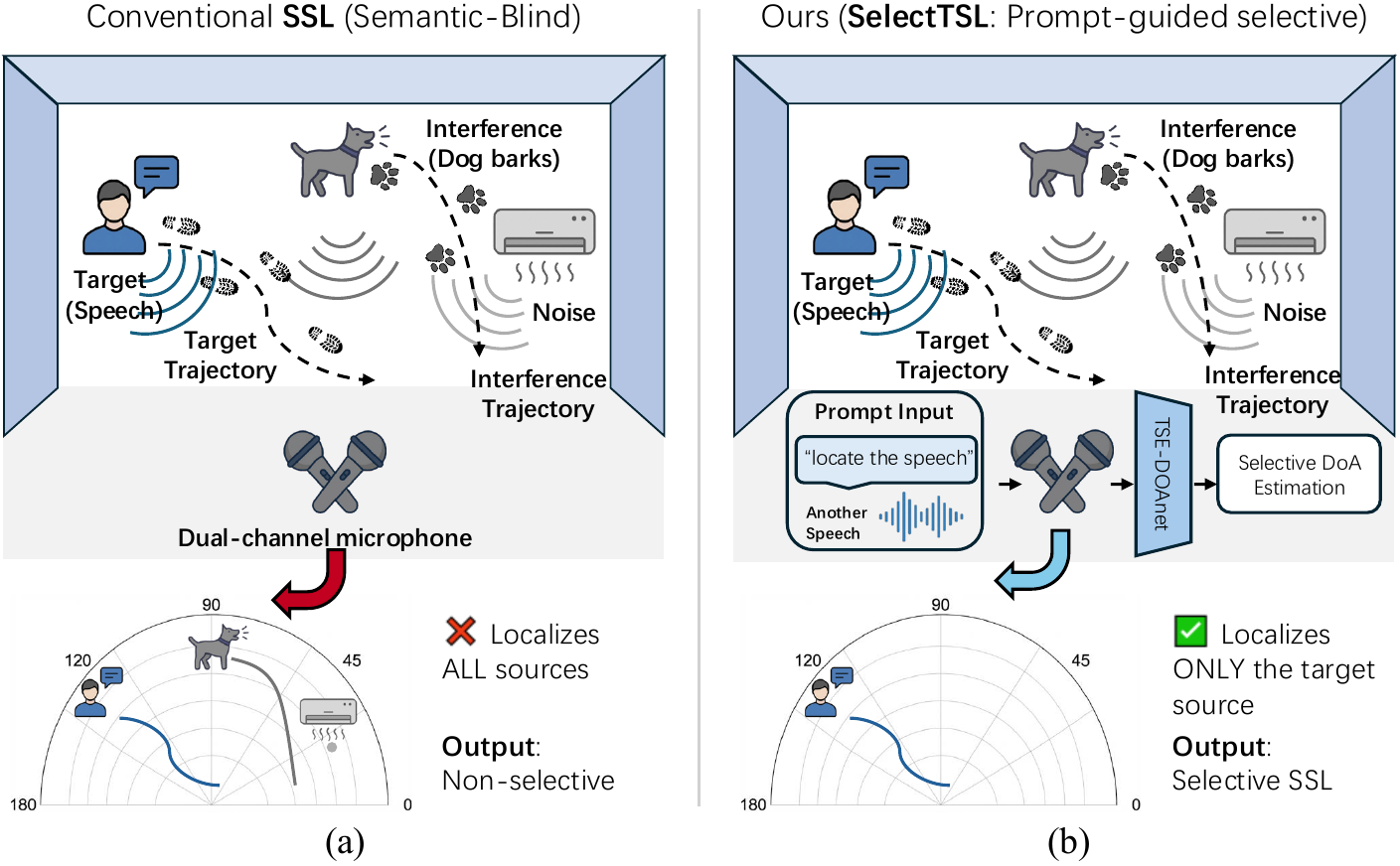}
    \caption{Illustration of conventional semantic-blind SSL and our proposed prompt-guided SelectTSL: (a) Conventional SSL localizes all active sources, yielding non-selective DoA trajectories, whereas (b) SelectTSL uses a text and audio prompt (``locate the speech'') to focus on the target speech which only provides its corresponding DoA trajectory.}
    \label{fig:intro_example}
\end{figure}

\IEEEPARstart{S}{ound} source localization (SSL), which estimates the spatial location of acoustic events, 
supports a wide range of array-based audio and speech applications.
For instance, in smart speakers, SSL enables direction-aware target speech enhancement and far-field automatic speech recognition (ASR) by steering microphone-array beamformers \cite{benesty2008microphone}.
In hearing aids, it supports directional noise reduction via spatial filtering \cite{heb2025microphone}.
Despite remarkable progress of SSL via signal processing (e.g., generalized cross-correlation with phase transform (GCC-PHAT)~\cite{knapp2003generalized}, multiple signal classification (MUSIC)~\cite{schmidt1986multiple}) or deep neural networks (e.g., convolutional recurrent neural networks (CRNNs)~\cite{chakrabarty2019multi}), existing methods are inherently \textit{semantic-blind}, treating all acoustic sources as generic signals without identity awareness: all active sources are localized without selectivity, including interfering sounds.
This differs from human auditory perception where listeners can selectively attend to a target sound when competing speakers and background noise exist, i.e., the ``cocktail party problem'' \cite{cherry1953some}.
As illustrated in Fig.~\ref{fig:intro_example}(a), conventional SSL systems indiscriminately localize both target and irrelevant sources (e.g., dog barks and noise), yielding non-selective direction-of-arrival (DoA) trajectories and preventing users from directing the system to focus on a specific target such as speech.

To associate semantics with location, sound event localization and detection (SELD) methods~\cite{adavanne2018sound, adavanne2018sound, politis2020overview} detect and localize all events of known classes. However, they perform passive scene analysis that lacks an interactive mechanism to filter sources based on user intent~\cite{senocak2025toward}. Conversely, in the field of target sound extraction (TSE), methods~\cite{wang2019voicefilter, liu2022separate, saijo2025leveraging, kim2025contextual} extract specific sound sources guided by audio or text prompts. Nevertheless, TSE focuses more on waveform reconstruction process rather than spatial sensing,
thus spatial cues like inter-channel phase difference (IPD) are often degraded for precise SSL.
This leads to a fundamental mismatch: TSE is aware of what to extract but not where to localize, whereas conventional SSL knows where sources are but not which semantic target to follow. Therefore, it remains an open problem to build a unified framework that is both semantic-aware and location-aware.


In this paper, we propose the SelectTSL network.
Unlike standard approaches, SelectTSL is an end-to-end localization framework, where prompt-guided target extraction guides subsequent spatial estimation.
In particular, a Prompt-Guided Selective Attention (PGSA) module, conditioned by multimodal prompts (text or audio), acts as a semantic filter to purify target-consistent features from the audio mixture. These features are then fused with spatial cues (e.g., IPD) and fed into a dedicated DoA estimator. This design decouples semantic selection from spatial estimation, enabling robust localization of user-specified targets even in noisy and multi-source environments.
Our contributions are listed as:



\begin{enumerate}
    \item We formulate a prompt-guided selective target sound localization task in dynamic, noisy and multi-source environments, where sources are moving and the number of active target sources is unknown and varies over time. The goal is to localize only the user-specified target(s) while suppressing interference.

    \item We propose \textbf{SelectTSL}, an end-to-end framework that supports both text and audio prompts. A prompt-guided PGSA module produces extraction-informed embeddings (EIEs) that condition an IPD Enhancer to refine spatial phase cues, which are fused with target magnitudes for DoA estimation.

    \item We handle unknown and time-varying active targets via a lightweight cardinality head that jointly predicts DoA heatmaps and frame-level target-source counts, enabling stable localization and tracking under moving speakers and intermittent activity.

    \item Extensive experiments 
    on synthetic and real data
    demonstrate consistent improvements over strong baselines in localization and tracking, and confirm robust generalization under realistic room acoustics.

    
    
\end{enumerate}

\begin{table*}[!h]
\centering
\caption{Baseline taxonomy with \textbf{track-slot (track-wise)} priority. Categories marked\textsuperscript{\dag} are non--track-wise (no fixed-track outputs).}
\label{tab:baselines}
\small
\setlength{\tabcolsep}{4pt}
\renewcommand{\arraystretch}{1.08}
\begin{tabularx}{\textwidth}{@{} l L L L L @{}}
\toprule
\textbf{Category} & \textbf{Model} & \textbf{Task family} & \textbf{Input / Feature} & \textbf{Target} \\
\midrule
\multirow{11}{*}{\textbf{Track-wise}}
& Multi-ACCDoA           & SELD     & Mch.\ STFT                 & Multi-ACCDoA \\
& IPDNet                 & SSL      & Mch.\ STFT                 & DoA (via DP\mbox{-}IPD) \\
& MIMO-DoAnet            & SSL      & Mch.\ STFT                 & Multi-out DoA \\
& EINV2                  & SELD     & FOA                        & Track-wise SELD \\
& embed-ACCDoA           & SELD     & CLAP-aug.\ feat.           & ACCDoA \\
& DiffTrack-DoA          & SSL      & Mch.\ STFT                 & DoA + diff.\ track loss \\
& SALSA / SALSA-Lite     & SELD     & SALSA / SALSA-Lite         & DoA + act. \\
& SE-ResNet              & SELD     & ResNet/SE enc.             & DoA + act. \\
& NGCC-SELD              & SELD     & NGCC + SELD pipe.          & DoA + act. \\
& CST-Former             & SELD     & CST attention              & DoA + act. \\
& DCASE25 & SELD     & Stereo (2-ch) audio        & DoA + act. \\
\midrule
\multirow{3}{*}{\textbf{SELD+DoA}\textsuperscript{\dag}}
& SWG\_Former            & SELD     & Transformer mod.           & DoA + act. \\
& SELDnet                & SELD     & Mel/FOA                    & ACCDoA \\
& SELDT                  & SELD(T)  & Transformer SELD           & DoA + act. \\
\midrule
\multirow{3}{*}{\textbf{Pure DoA}\textsuperscript{\dag}}
& FN-SSL                 & SSL      & FB+NB fusion               & DoA (via DP\mbox{-}IPD) (1src) \\
& SRP-DNN                & SSL      & DP cues + SRP feat.        & SRP spectrum \\
& SALADnet               & SSL      & FOA                        & Attn.\ DoA \\
\midrule
\multirow{5}{*}{\textbf{Prompt-based}\textsuperscript{\dag}}
& KeywordLoc             & SSL (query)   & Mch.\ feat. + kw cue           & kw-cond.\ spk DoA \\
& Class-cond.\ SELD      & SELD (query)  & Mch.\ feat. + 1hot cls (FiLM)  & cls-cond.\ (DoA+act.) \\
& Text-Queried SEL       & SEL (query)   & GCC-PHAT + txt emb.            & txt-cond.\ DoA/traj \\
& LocSelect              & SSL (query)   & 2-ch spec. + enroll sp.        & enroll-cond.\ spk DoA \\
& GCC-Speaker            & SSL (query)   & GCC-PHAT + spk-wt              & spk DoA \\
\bottomrule
\end{tabularx}

\vspace{2pt}
\raggedright
\textbf{Abbrev.} Mch.=multichannel; FOA=First-order Ambisonics; FB/NB=full-/narrow-band; STFT=short-time Fourier transform; CLAP=Contrastive Language-Audio Pretraining; ACCDoA=activity-coupled Cartesian DoA; DP-IPD=direct-path inter-channel phase difference; SRP=steered response power; NGCC=neural generalized cross-correlation; SE=squeeze-and-excitation; enc.=encoder; feat.=feature; mod.=module; pipe.=pipeline; act.=activity; diff.=differentiable; aug.=augmented; spk=speaker; spec.=spectrogram; kw=keyword; cls=class; txt=text; traj=trajectory; 1hot=one-hot; wt=weighting; \textsuperscript{\dag}\,Non--track-wise (no fixed-track outputs).
\vspace{-4mm}
\end{table*}

\section{Related Work}
\label{sec:related_work}
Conventional SSL approaches excel at estimating spatial direction but are typically prompt-agnostic, localizing all active sources indiscriminately.
Conversely, TSE provides controllable target selection via auxiliary cues to reconstruct clean sound. However, it prioritizes signal fidelity over spatial consistency, often distorting the spatial information needed for precise localization.
Prompt-guided selective target sound localization bridges the gap between semantic-aware extraction and spatial-aware localization, effectively unifying these two objectives.
Despite its promise, existing studies in this emerging field are still limited by unimodal prompting (text or audio only) and restricted evaluation scenarios (e.g., handling at most a single target per query, with limited evaluation in moving-source and noisy environments).
The following section reviews literature in TSE, SSL, and prompt-guided localization.
The significant methods summarized in Table~\ref{tab:baselines}.

\subsection{Target Sound Extraction (TSE)}
TSE aims to extract a specific type of sound source (e.g., speech, music, or other acoustic events) from a mixture, conditioned on explicit auxiliary cues.
Existing methods have explored auxiliary cues ranging from visual signals~\cite{afouras2018deep,li2024audio,pan2022usev,tao2025seanet}, spatial information~\cite{gu2019neural,gu2022towards}, reference audio~\cite{wang2019voicefilter,vzmolikova2019speakerbeam,xu2020spex,liu2023x} and semantic text descriptions~\cite{liu2022separate,dong2022clipsep,wu2023large,ma2024clapsep}. 
While visual cues are highly effective in speech-oriented settings by leveraging articulatory motions (e.g., lip/face movements)~\cite{afouras2018deep,li2024audio}, they are not generally applicable to arbitrary acoustic events. Spatial cues used as auxiliary prompts often require additional priors (e.g., array geometry or target direction) beyond the mixture audio, and are thus beyond our scope.
Therefore, we focus on text and audio cues in the remainder of this section.

For text-prompted extraction, the core mechanism involves mapping a natural language caption (e.g., ``piano'', ``glass breaking'') into a semantic embedding space, utilizing pre-trained contrastive models such as Contrastive Language-Audio Pretraining (CLAP)~\cite{wu2023large}.
This embedding then acts as a condition to steer the separation network, predicting a time-frequency mask or waveform residual that extracts the source matching the description~\cite{liu2022separate,liu2024separate,ma2024clapsep}.

In contrast, audio-prompted extraction utilizes a reference audio clip as the enrollment cue.
This mechanism generally operates in two distinct modes based on the granularity of the guidance:
(i) query by example, where the reference shares the same semantic class as the target but differs in instance (e.g., using a generic dog bark to extract a specific dog)~\cite{chen2022zero}; and
(ii) target enrollment, commonly used in TSE, which utilizes a clean reference utterance to guide the extraction of the target speaker~\cite{wang2019voicefilter,vzmolikova2019speakerbeam,liu2023x}.

Architecturally, these prompt encoders are integrated into separation backbones, such as Conv-TasNet~\cite{luo2019conv}, Dual-Path Recurrent Neural Network (DPRNN)~\cite{luo2020dual} and TF-GridNet~\cite{wang2023tf}, via varying fusion mechanisms, ranging from simple feature-wise linear modulation (FiLM) to more complex cross-attention layers.
However, while prompt-guided TSE offers fine-grained controllability over which source to recover, it is primarily optimized for signal reconstruction fidelity (e.g., scale-invariant signal-to-noise ratio (SI-SNR)) rather than spatial consistency.
Unless explicitly designed to preserve inter-channel phase consistency, the reconstruction objective of TSE tends to distort the spatial cues essential for localization.
In essence, TSE determines what to extract, yet remains agnostic to where the source is located.


\subsection{Sound Source Localization (SSL)}
In contrast to TSE that controls which source to extract, SSL addresses where sources are by estimating DoA from multichannel audio inputs~\cite{grumiaux2022survey}. Recent advances have largely adopted deep learning to map time-frequency (TF) and spatial features to DoA estimates. 
In parallel, the vision community has studied audio-visual sound source localization in videos, where visual context is used to select and localize the sounding object~\cite{senocak2019learning, xuan2024robust, song2024enhancing}.

Contemporary systems generally follow a common pipeline: a feature extraction backbone followed by a task-specific head. Despite this shared framework, these methods differ primarily in two aspects: (i) the choice of input representations, ranging from raw short-time Fourier transform (STFT) spectrograms to spatially explicit cues such as IPD, inter-channel level difference (ILD), covariance eigenvectors, and generalized cross correlation (GCC)~\cite{adavanne2018sound,nguyen2022salsa}; and (ii) the modeling strategy for these representations, spanning from bandwise processing that preserves narrow-band spatial cues to full-band mechanisms that capture cross-frequency correlations~\cite{wang2023fn,wang2024ipdnet}. We organize this subsection along these two aspects to explain the sources of recent performance gains.

Typical SSL models map TF features to DoA via classification or regression heads using deep neural backbones, such as convolutional neural networks (CNNs), CRNNs, Conformer, and Transformer~\cite{adavanne2018sound,grumiaux2021saladnet,huang2023swg,hu2022sound,shul2024cst}. Beyond raw STFT features, many systems explicitly encode spatial cues. For example, SALSA augments log-spectrograms with the normalized principal eigenvector of the spatial covariance~\cite{nguyen2022salsa}, and SALSA-Lite uses normalized IPD~\cite{nguyen2022salsa_lite}. Spatial correlation features such as GCC-PHAT and its neural variant neural GCC-PHAT (NGCC-PHAT) are also commonly used in recent SSL systems~\cite{knapp2003generalized,berg2024lu,berg2024learning}. 
In addition to refined inputs, recent work improves robustness by modeling direct-path cues and cross-frequency structure. For example, FN-SSL and IPDNet use full- and narrow-band fusion to estimate direct-path inter-channel phase difference (DP-IPD) and infer DoA (e.g., via template matching)~\cite{wang2023fn,wang2024ipdnet}. SRP-DNN learns direct-path delays and phase, and injects them into steered response power (SRP)-style spectra to refine spatial peaks~\cite{yang2022srp}. Moreover, MIMO-DoAnet predicts source-wise pseudo-spectra to reduce reliance on post-hoc heuristics~\cite{yin2022mimo}.

Beyond DoA estimation, SELD extends the SSL objective by jointly estimating the sound event type. In this case, SELD models can be used as SSL baselines by solely using its localization branch. For instance, ACCDoA and Multi-ACCDoA~\cite{shimada2021accdoa,shimada2022multi} use activity-coupled Cartesian DoA (ACCDoA) vectors as regression targets, with auxiliary duplicating permutation invariant training (ADPIT) to handle sound overlaps of the same class.
Another line of work adopts explicit track slots with permutation-invariant training. Representative methods such as EIN~\cite{cao2020event} employ separate attention-based tracks and have been widely adopted in challenge systems~\cite{cao2021improved,hu2022sound}. Variants mainly differ in encoders/backbones (e.g., SALADnet~\cite{grumiaux2021saladnet}, SwG-former~\cite{huang2023swg}, CST-Former~\cite{shul2024cst}) and feature fusion strategies~\cite{mu2024mff}. The Detection and Classification of Acoustic Scenes and Events (DCASE) 2025 Task 3 baseline further shows an audio-only stereo setting with a CRNN and a Multi-ACCDoA-like output~\cite{shimada2025stereo}.

While the aforementioned methods focus on frame-level DoA estimates, moving-source scenarios require associating predictions over time. For example, SELDT~\cite{adavanne2019localization} forms trajectories by linking frame-wise source activity and DoA estimates. Differentiable tracking~\cite{adavanne2021differentiable} further integrates identity assignment (e.g., Hungarian matching) into training to optimize sequence-level objectives. Although effective for trajectory estimation, such track-based designs inherently rely on a fixed number of output slots and lack a mechanism for user-specified target selection. 
This limitation motivates prompt-guided localization reviewed below.

\subsection{Prompt-guided Localization}
Prompt-guided localization conditions a DoA estimator on a user-specified cue, enabling it to localize the queried target selectively rather than all active sources indiscriminately. Existing work has explored this idea through two modalities: text-based conditioning for semantic descriptions and audio-based conditioning for acoustic enrollment.

On the text side, early work already explored text-like cues in the form of fixed keywords.
Keyword-based speaker localization~\cite{sivasankaran2018keyword} introduces the task of localizing the speaker who uttered a predefined trigger phrase (e.g., a wake word) in the presence of overlapping speech. 
Beyond fixed keywords, class-conditioned localization provides a related formulation where the query is a target event class rather than free-form text.
A class-conditioned SELD framework~\cite{slizovskaia2022locate} utilizes one-hot class indicators as conditioning cues, injecting them into an ACCDoA-based baseline via FiLM layers. This design enables the model to selectively estimate the DoA of the queried class while treating concurrent events from other classes as interference during training.
More recently, SEL~\cite{zhao2024text} extends text conditioning to free-form captions and benchmarks fusion schemes that combine textual embeddings (e.g., CLAP, BERT, or FlanT5) with spatial audio features such as GCC-PHAT for selective DoA prediction and trajectory estimation.

For audio-based localization, LocSelect~\cite{chen2024locselect} studies target speaker localization given an enrollment utterance of the same target speaker. It first predicts a speaker-dependent spectrogram mask to suppress interferers and then estimates the target DoA from the filtered spectrogram using a long short-term memory (LSTM)-based localizer~\cite{chen2024locselect}.
Alternatively, approaches like GCC-Speaker~\cite{li2023gcc} adapt classical spatial features directly, learning speaker-dependent weights for GCC-PHAT to enhance selectivity in multi-speaker scenarios.

Overall, existing prompt-guided localization studies are predominantly unimodal, utilizing either text or audio cues but not both within a unified model.
Moreover, evaluations are often confined to simplified settings, with limited coverage of moving targets, noisy mixtures, or prompts that correspond to multiple targets.

\vspace{-3mm}

\section{Task Formulation}


Let us consider a dual-channel recording setup. 
The signal received at the $m$-th microphone ($m \in {1,2}$), denoted by $x_m(t)$, is modeled as the sum of $J$ source signals convolved with their respective room impulse responses (RIRs), together with additive ambient noise:
\begin{equation}
    x_m(t) = \sum_{j=1}^{J(t)} s_j(t) * h_{m,j}(t, \theta_j(t)) + n_m(t),
    \label{eq:signal_model}
\end{equation}
where $s_j(t)$ is the $j$-th source signal, $J(t)$ is the number of active sources at time $t$, $n_m(t)$ is the additive noise at the $m$-th microphone, and $h_{m,j}(t, \theta_j(t))$ represents the RIR between the $j$-th source and the $m$-th microphone, which is critically dependent on the time-varying DoA of the source $\theta_j(t)$. $\star$ indicates convolution.

Our goal is to learn a prompt-guided localization model $\mathcal{M}_{\psi}$, parameterized by $\psi$, designed to estimate the spatial trajectory of a target source specified by a multimodal prompt. 
Unlike traditional SSL systems, the proposed SelectTSL is conditioned on a user-provided cue. It maps the dual-channel mixture signal $\mathbf{x}(t) = [x_1(t), x_2(t)]^T$, together with an auxiliary audio cue $\mathbf{x}_{\text{cue}}(t)$ or a textual prompt $\mathbf{x}_{\text{text}}$, to two corresponding frame-level predictions:
\begin{itemize}
    \item \textbf{The DoA Posteriorgram}, $\hat{\mathbf{P}}_{\mathrm{DoA}} \in [0, 1]^{T_{\mathrm{out}} \times \Theta}$ denotes frame-level DoA confidence maps, where $T_{\mathrm{out}}$ is the number of output frames after temporal alignment and $\Theta=180$ is the number of azimuth bins discretized from $0^\circ$ to $179^\circ$ at $1^\circ$ resolution. Due to front-back ambiguity in symmetric dual-microphone arrays, predictions are restricted to a $180^\circ$ range, and the multi-label formulation allows multiple active bins per frame with continuous confidence values.

    \item \textbf{The Source Cardinality}, $\hat{\mathbf{P}}_{\mathrm{card}} \in [0, 1]^{T_{\mathrm{out}} \times 3}$ denotes frame-level cardinality predictions, a probability distribution over the presence of 0, 1, or 2 active target sources.
\end{itemize}
We optimize the model parameters $\psi$ to jointly predict the target sources’ DoA posteriorgram $\hat{\mathbf{P}}_{\mathrm{DoA}}$ and frame-level cardinality $\hat{\mathbf{P}}_{\mathrm{card}}$, as formally defined by:
\begin{equation}
    (\hat{\mathbf{P}}_{\mathrm{DoA}}, \hat{\mathbf{P}}_{\mathrm{card}}) = \mathcal{M}_{\psi}(\mathbf{x}(t)| \mathbf{x}_{\text{cue}}(t), \mathbf{x}_{\text{text}}).
\end{equation}

\section{SelectTSL Architecture}
\begin{figure*}[t]
    \centering
    \includegraphics[width=1.0\linewidth]{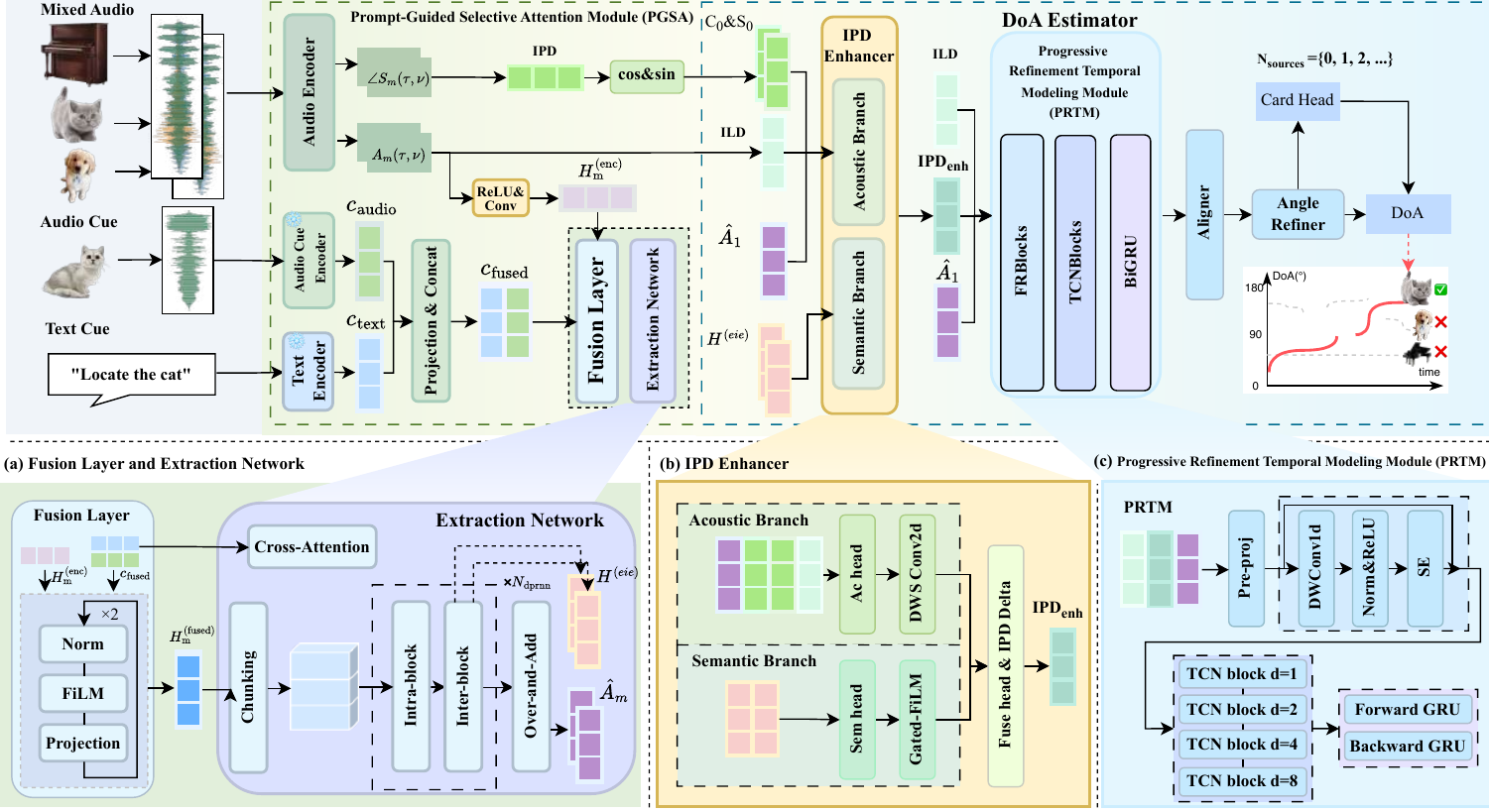}
    \caption{Overall architecture of SelectTSL. The PGSA module outputs the target magnitude spectrogram and a stack of EIEs $\mathbf{H}^{(\mathrm{eie})}$ (from the Extraction Network). The DoA estimator enhances spatial cues with these EIEs and aggregates semantic \& spatial information for localization and cardinality estimation.}
    \label{fig:architecture}
    
\end{figure*}

SelectTSL is an end-to-end framework for prompt-guided selective target sound localization, in which a PGSA module generates extraction-informed embeddings from a dual-channel mixture and a DoA estimator jointly predicts frame-level DoA and source cardinality. The overall architecture is shown in Fig.~\ref{fig:architecture} and is described as follows.

\subsection{Prompt-Guided Selective Attention Module (PGSA)}
\label{sec:pgsa}
We introduce the PGSA module, which acts as a prompt-guided selective filter. It leverages multimodal prompts (text or audio) to extract target-specific representations from the input mixture.


\subsubsection{Audio Encoder}
Given a dual-channel signal $\mathbf{x}_m(t)$ for $m\in\{1,2\}$, we compute complex spectrograms $\mathbf{S}_m(\tau,\nu)$ by short-time Fourier transform (STFT) using a Hann window, where $\tau \in \{1,\dots,T_{\mathrm{in}}\}$ and $\nu \in \{1,\dots,F\}$ represent time frames and frequency bins, respectively. We extract
\begin{itemize}
    \item \textbf{Magnitude cues:} $\mathbf{A}_m(\tau,\nu)=|\mathbf{S}_m(\tau,\nu)|$, which are encoded by a $\mathrm{Conv1D}$ followed by a $\mathrm{ReLU}$ module into
    $\mathbf{H}_m^{(\mathrm{enc})}\in\mathbb{R}^{D_{\mathrm{emb}}\times T_{\mathrm{in}}}$:
    \begin{equation}
        \mathbf{H}_m^{(\mathrm{enc})}=\mathrm{ReLU}(\mathrm{Conv1D}(\mathbf{A}_m)).
    \end{equation}
    \item \textbf{Spatial cues:} IPD and inter-channel level difference (ILD):
    \begin{align}
    \mathrm{IPD}(\tau,\nu) &= \angle \mathbf{S}_1(\tau,\nu)-\angle \mathbf{S}_2(\tau,\nu), \\
    \mathrm{ILD}(\tau,\nu) &= \log\!\frac{|\mathbf{S}_1(\tau,\nu)|+\epsilon}{|\mathbf{S}_2(\tau,\nu)|+\epsilon}.
    \end{align}

\end{itemize}

\subsubsection{Prompt Encoders}
We employ a frozen CLAP model to extract audio cue $\mathbf{x}_{\text{cue}}(t)$ and text $\mathbf{x}_{\text{text}}$:
\begin{align}
    \mathbf{c}_{\text{audio}} &= \mathrm{MLP}_{\text{cue}}(\mathrm{CLAP}_{\text{audio}}(\mathbf{x}_{\text{cue}}(t))) \in \mathbb{R}^{256},\\
    \mathbf{c}_{\text{text}}  &= \mathrm{MLP}_{\text{text}}(\mathrm{CLAP}_{\text{text}}(\mathbf{x}_{\text{text}})) \in \mathbb{R}^{256}.
\end{align}
Here, $\mathrm{MLP}_{\text{cue}}$ and $\mathrm{MLP}_{\text{text}}$ serve as projection layers. During training, we freeze the CLAP encoders and optimize only the projection layers.
We then form a unified guidance vector $\mathbf{c}_{\text{fused}}=[\mathbf{c}_{\text{audio}};\mathbf{c}_{\text{text}}]\in\mathbb{R}^{512}$.
In our model, the audio prompt $\mathbf{x}_{\text{cue}}(t)$ is taken as the first $1\,\mathrm{s}$ of a $6\,\mathrm{s}$ clip, and the remaining $5\,\mathrm{s}$ of that clip is used as the target segment for selection and localization.

\subsubsection{Fusion Layer}
We condition the encoded magnitude features $\mathbf{H}_m^{(\text{enc})}$ using a two-stage FiLM cascade driven by $\mathbf{c}_{\text{fused}}$:
\begin{equation}
    \mathrm{FiLM}(\mathbf{h},\mathbf{c})=\gamma(\mathbf{\mathbf{c}_{\text{fused}}})\odot \mathbf{h}+\beta(\mathbf{\mathbf{c}_{\text{fused}}}),
\end{equation}
where $\gamma(\cdot)$ and $\beta(\cdot)$ are learnable functions that map the conditioning vector to scale and shift parameters.
First, $\mathbf{c}_{\text{fused}}$ modulates the full-dimensional $\mathbf{H}_m^{(\text{enc})}$, followed by a projection. Then, a downsampled guidance modulates the projected features, yielding $\mathbf{H}_m^{(\text{fused})}$.

\subsubsection{Extraction Network}
We segment $\mathbf{H}_m^{(\text{fused})}$ into overlapping chunks of length $L_{\mathrm{chunk}}$ with 50\% overlap, and process them using $N_{\mathrm{dprnn}}$ dual-path blocks (intra-/inter-chunk RNNs from DPRNN~\cite{luo2020dual}).
Inside the stack, we periodically inject cross-attention between the current audio features and the semantic guidance $\mathbf{c}_{\text{fused}}$. 
Let $\mathbf{H}\in\mathbb{R}^{T\times D_h}$ denote the frame-level features in a block and $\mathbf{c}_{\text{fused}}\in\mathbb{R}^{D_c}$ the guidance vector. We broadcast the guidance across time as $\tilde{\mathbf{C}}\in\mathbb{R}^{T\times D_c}$ and compute cross-attention:
\begin{align}
    \mathbf{Q} &= \tilde{\mathbf{C}}\mathbf{W}^{\mathrm{Q}}, &
    \mathbf{K} &= \mathbf{H}\mathbf{W}^{\mathrm{K}}, &
    \mathbf{V} &= \mathbf{H}\mathbf{W}^{\mathrm{V}},
\end{align}
with $\mathbf{W}^{\mathrm{Q}}\!\in\!\mathbb{R}^{D_c\times d_k}$ and $\mathbf{W}^{\mathrm{K}},\mathbf{W}^{\mathrm{V}}\!\in\!\mathbb{R}^{D_h\times d_k}$. The attention weights and residual update are
\begin{align}
    \boldsymbol{\alpha} &= \mathrm{softmax}\!\left(\frac{\mathbf{Q}\mathbf{K}^\top}{\sqrt{d_k}}\right), &
    \mathbf{H} &\leftarrow \mathbf{H} + \boldsymbol{\alpha}\mathbf{V}.
\end{align}

After the DPRNN stack, we project its output to a soft mask $\mathbf{M}_m$, which filters the encoder features and a decoder reconstructs the target magnitude:
\begin{equation}
    \hat{\mathbf{H}}_{m}=\mathbf{M}_m\odot \mathbf{H}_m^{(\text{enc})},\qquad
    \hat{\mathbf{A}}_{m}=\mathrm{MLP}_{\text{dec}}(\hat{\mathbf{H}}_{m}).
\end{equation}
We use the extracted target features $\hat{\mathbf{A}}_1\in\mathbb{R}^{F\times T}$ for DoA estimation, where $F$ and $T$ denote frequency bins and frames.

From the last $K_{\mathrm{EIE}}$ DPRNN blocks (before overlap-and-add), we export extraction-informed embeddings (EIEs). Each block’s features are linearly projected to a TF feature map in $\mathbb{R}^{F\times T_{\mathrm{in}}}$, as illustrated in the Extraction Network (bottom-left) of Fig.~\ref{fig:architecture}.
Averaging across the two channels yields $\mathbf{H}^{(\mathrm{eie})}_k\in\mathbb{R}^{F\times T_{\mathrm{in}}}$, $k=1,\dots,K_{\mathrm{EIE}}$. Stacking forms
\begin{equation}
\mathbf{H}^{(\mathrm{eie})}
=\mathrm{stack}\big(\mathbf{H}^{(\mathrm{eie})}_1,\dots,\mathbf{H}^{(\mathrm{eie})}_{K_{\mathrm{EIE}}}\big)
\in\mathbb{R}^{K_{\mathrm{EIE}}\times F\times T_{\mathrm{in}}}.
\end{equation}
These maps emphasize TF regions that are relevant for extracting the target and serve as guidance for IPD enhancement.

\subsection{DoA Estimator}
\label{sec:DoA_estimator}
The DoA estimator consists of two modules: an IPD Enhancer that refines spatial cues, and a Progressive Refinement Temporal Modeling Module (PRTM).
Although the PGSA module outputs $\hat{\mathbf{A}}_m$ for each channel $m$, these per-channel magnitude features are highly redundant. Therefore, we use only $\hat{\mathbf{A}}_1$ as the magnitude input to the DoA estimator.
The IPD Enhancer takes the mixture spatial cues (IPD, ILD), the extracted target features $\hat{\mathbf{A}}_1$ from the PGSA module, and the $\mathbf{H}^{(\mathrm{eie})}$ from the Extraction Network. It then outputs the enhanced IPD $\mathrm{IPD}_{\mathrm{enh}}$.
The PRTM operates on $\hat{\mathbf{A}}_1$, $\mathrm{IPD}_{\mathrm{enh}}$, and ILD to produce (i) a frame-level DoA posteriorgram over $\Theta$ azimuth bins and (ii) a categorical distribution over the source-count set $\mathcal{N}=\{0,1,\dots,N_{\max}\}$ (with $N_{\max}=2$ in our experiments), where $N_{\max}$ is the maximum number of simultaneously active target sources.


\subsubsection{IPD Enhancer}
We refine the mixture IPD with an IPD Enhancer that is explicitly conditioned on both low-level spatial cues and high-level EIEs (Fig.~\ref{fig:architecture}).

\textbf{Acoustic branch.} The input to the acoustic branch comprises $\hat{\mathbf{A}}_1$, $\mathbf{C}_0=\cos(\mathrm{IPD})$, $\mathbf{S}_0=\sin(\mathrm{IPD})$, and $\mathrm{ILD}$. These features are concatenated to form a 4-channel spectrogram representation, which is subsequently processed by an acoustic feature encoder (denoted as \textit{ac head}). This module employs a lightweight depthwise-separable Conv2d (DWS Conv2d) to efficiently capture fine-grained spectro-spatial features.

\textbf{Semantic branch.} Operating in parallel, this branch exclusively processes $\mathbf{H}^{(\mathrm{eie})}$ from the Extraction Network. It is encoded by a semantic head into a conditioning representation that drives a gated FiLM modulation. Specifically, global pooling is employed to derive channel-wise scale and shift parameters, while a concurrent spatial gate modulates the acoustic features to enable spatially selective IPD enhancement.

\textbf{Fusion and residual prediction.}
Let $\mathbf{A}_{\mathrm{ac}}$ denote the acoustic features produced by the acoustic branch and $\mathbf{A}_{\mathrm{sem}}$ the semantic features produced by the semantic head.
We fuse them by applying FiLM conditioning followed by a gated modulation:
\begin{equation}
\tilde{\mathbf{A}}
= \big(\mathbf{1}+g(\mathbf{A}_{\mathrm{sem}})\big)\odot \mathrm{FiLM}(\mathbf{A}_{\mathrm{ac}}, \mathbf{A}_{\mathrm{sem}}),
\label{eq:gated_film}
\end{equation}
where $g(\cdot)$ is a spatial gating head that modulates the FiLM-conditioned features to enable spatially selective IPD enhancement.
A fuse head then predicts cosine--sine residuals (IPD deltas), and the enhanced IPD is recovered by the four-quadrant inverse tangent $\operatorname{atan2}(\cdot,\cdot)$:
\begin{equation}
\begin{aligned}
\mathbf{C}_0 &= \cos(\mathrm{IPD}), \qquad \mathbf{S}_0 = \sin(\mathrm{IPD}),\\
&\hfill (\Delta\mathbf{C},\Delta\mathbf{S}) = \Delta_{\mathrm{IPD}}(\tilde{\mathbf{A}}), \hfill\\
\mathrm{IPD}_{\mathrm{enh}} &= \operatorname{atan2}\!\big(\mathbf{S}_0+\Delta\mathbf{S},\,\mathbf{C}_0+\Delta\mathbf{C}\big)
\in \mathbb{R}^{F\times T_{\mathrm{in}}}.
\end{aligned}
\label{eq:ipd_enh}
\end{equation}


\subsubsection{Progressive Refinement and Temporal Modeling (PRTM) Module}
For each frame $\tau$, we concatenate the features:
\begin{equation}
    \mathbf{z}_\tau
    = \big[\hat{\mathbf{A}}_1(:,\tau);\;\mathrm{IPD}_{\text{enh}}(:,\tau);\;\text{ILD}(:,\tau)\big].
\end{equation}
Stacking over time gives $\mathbf{Z}_{\text{in}}\in\mathbb{R}^{T_{\mathrm{in}}\times C_{\mathrm{feat}}}$. The sequence is then processed by a PRTM module. 
First, the input is projected and processed by a stack of Feature Refinement Blocks (FRBs), whose composite operation is denoted as $\mathcal{R}(\cdot)$. 
This module progressively fuses three distinct cues: target magnitude, enhanced IPD, and ILD. 
Each FRB is implemented using depthwise-separable 1D convolutions, squeeze-and-excitation modules, and residual connections.
This progressive fusion projects the inputs into a unified feature space where spatial and spectral evidence are tightly coupled. On top of this fused representation, a dilated temporal convolutional network (TCN) and a bidirectional gated recurrent unit (BiGRU) focus on modeling longer-range temporal patterns:
\begin{equation}
    \mathbf{H}^{(\text{temp})}
    = \mathrm{BiGRU}\!\left(\mathrm{TCN}\big(\mathcal{R}(\mathrm{Linear}(\mathbf{Z}_{\text{in}}))\big)\right)
    \in\mathbb{R}^{T_{\mathrm{in}}\times 2D_{\mathrm{gru}}}.
\end{equation}

\noindent A temporal alignment layer adjusts the frame rate to obtain
\begin{equation}
    \mathbf{H}^{(\text{aligned})}
    = \mathcal{T}\!\left(\mathbf{H}^{(\text{temp})}\right)
    \in\mathbb{R}^{T_{\mathrm{out}}\times 2D_{\mathrm{gru}}}.
\end{equation}
The alignment layer $\mathcal{T}(\cdot)$ compresses the sequence to a fixed length $T_{\mathrm{out}}$, so that the prediction heads operate at the same temporal resolution as the labels.

\subsubsection{Prediction Heads}
The aligned features $\mathbf{H}^{(\mathrm{aligned})}\!\in\!\mathbb{R}^{T_{\mathrm{out}}\times 2D_{\mathrm{gru}}}$
are decoded by the Angle Refiner (Fig.~\ref{fig:architecture}), which is implemented as two parallel heads. Let $\mathrm{proj}_{\mathrm{DoA}}(\cdot)$ and
$\mathrm{proj}_{\mathrm{card}}(\cdot)$ denote two learned frame-wise affine projections that output logits, applied row-wise to $\mathbf{H}^{(\mathrm{aligned})}$:
\begin{subequations}
\label{eq:heads}
\begin{align}
\hat{\mathbf{P}}_{\mathrm{DoA}}
&= \sigma\!\Big(\mathrm{proj}_{\mathrm{DoA}}\big(\mathbf{H}^{(\mathrm{aligned})}\big)\Big)
\in [0,1]^{T_{\mathrm{out}}\times \Theta},
\label{eq:DoA_head} \\
\hat{\mathbf{P}}_{\mathrm{card}}
&= \mathrm{softmax}\!\Big(\mathrm{proj}_{\mathrm{card}}\big(\mathbf{H}^{(\mathrm{aligned})}\big)\Big)
\in [0,1]^{T_{\mathrm{out}}\times |\mathcal{N}|},
\label{eq:card_head}
\end{align}
\end{subequations}
where $\mathrm{proj}_{\mathrm{DoA}}:\mathbb{R}^{2D_{\mathrm{gru}}}\!\to\!\mathbb{R}^{\Theta}$ and
$\mathrm{proj}_{\mathrm{card}}:\mathbb{R}^{2D_{\mathrm{gru}}}\!\to\!\mathbb{R}^{|\mathcal{N}|}$.

During inference, we strictly couple the two predictions frame-wise. 
First, the number of active sources is estimated as $\hat{n}_t=\arg\max_{n\in\mathcal{N}}\hat{\mathbf{P}}_{\mathrm{card}}(t,n)$. 
Then, the DoA estimates are derived from the top-$\hat{n}_t$ peaks of $\hat{\mathbf{P}}_{\mathrm{DoA}}(t,:)$. 
Specifically, peak picking involves: 
(i) circular Gaussian smoothing on $\hat{\mathbf{P}}_{\mathrm{DoA}}(t,:)$ to mitigate jaggedness while handling the $0^\circ$-$180^\circ$ wrap-around; 
(ii) identifying strict local maxima that exceed an adaptive threshold $\tau_t=\max(\mu_t+\sigma_t,\tau_{\min})$ (where $\mu_t,\sigma_t$ are the frame-wise mean and std, and $\tau_{\min}=0.3$); 
and (iii) selecting the $\hat{n}_t$ highest peaks as the final DoA predictions.

\subsubsection{Training Objective}
The network is jointly trained with three loss terms. The separation loss is the negative SI-SNR averaged over the two channels:
\begin{equation}
    \mathcal{L}_{\mathrm{sel}}
    = -\frac{1}{2}\Big(
        \mathrm{SI\text{-}SNR}(\hat{\mathbf{s}}_1,\mathbf{s}_1)
      + \mathrm{SI\text{-}SNR}(\hat{\mathbf{s}}_2,\mathbf{s}_2)
      \Big).
\end{equation}
For localization, we use a Binary Cross Entropy (BCE) for frame-level DoA estimation and a Cross Entropy (CE) for source-count prediction:
\begin{equation}
    \mathcal{L}_{\mathrm{DoA}}
    = \mathrm{BCE}\big(\hat{\mathbf{P}}_{\mathrm{DoA}},\mathbf{P}^{\star}_{\mathrm{DoA}}\big),\quad
    \mathcal{L}_{\mathrm{card}}
    = \mathrm{CE}\big(\hat{\mathbf{P}}_{\mathrm{card}},\mathbf{n}^{\star}\big),
\end{equation}
where $\mathbf{P}^{\star}_{\mathrm{DoA}}\in[0,1]^{T_{\mathrm{out}}\times\Theta}$ is a soft DoA label built from ground-truth (GT) azimuths, and $\mathbf{n}^{\star}\in\{0,1,2\}^{T_{\mathrm{out}}}$ is the frame-wise source-count label.

The total loss is a weighted sum of separation, DoA and cardinality terms:
\begin{equation}
    \mathcal{L}
    = \omega_{\mathrm{sel}}\,\mathcal{L}_{\mathrm{sel}}
      + \omega_{\mathrm{DoA}}\,\mathcal{L}_{\mathrm{DoA}}
      + \omega_{\mathrm{card}}\,\mathcal{L}_{\mathrm{card}},
\end{equation}
where we fix $\omega_{\mathrm{sel}}=1$ and tune the other weights to balance the influence of each task's gradient. This is crucial as the losses operate at different scales, especially near convergence.

The selection loss $\mathcal{L}_{\mathrm{sel}}$ (negative SI-SNR) typically converges to a magnitude on the order of $O(10^1)$.
In contrast, $\mathcal{L}_{\mathrm{DoA}}$ is a BCE averaged over $T_{\mathrm{out}} \times \Theta$ bins. Due to the sparse nature of the DoA target (at most 2 of 180 bins are active), a converging model predicts near-zero loss for $\approx 99\%$ of the outputs. Consequently, the average loss $\mathcal{L}_{\mathrm{DoA}}$ becomes very small, e.g., $O(10^{-1})$ or $O(10^{-2})$.
To prevent the DoA gradient from vanishing and ensure its influence remains comparable to the separation term (i.e., $\omega_{\mathrm{DoA}}\mathcal{L}_{\mathrm{DoA}} \approx \omega_{\mathrm{sel}}\mathcal{L}_{\mathrm{sel}}$), we need to compensate for this scale mismatch. Therefore, $\omega_{\mathrm{DoA}}$ is chosen to be on the order of $O(10^2)$.

\vspace{-4mm}
\section{Experimental setting}

\subsection{Dataset}
\begin{table}[t]
\centering
\caption{Dataset statistics and key settings.}
\label{tab:data_stats}
\small
\setlength{\tabcolsep}{4pt}
\renewcommand{\arraystretch}{1.08}
\begin{tabularx}{\columnwidth}{@{}p{0.47\columnwidth} >{\raggedright\arraybackslash}X@{}}
\toprule
\textbf{Item} & \textbf{Value} \\
\midrule
Simulated dataset (tr/val/tt) &
\makecell[l]{Hours: 288.9 / 31.1 / 18.1\\
Clips: 208,008 / 22,392 / 13,032} \\

Real dataset (tr/val/tt) &
\makecell[l]{TAU-SRIR: 9 rooms; \\
Clips: 72,000 / 9,000 / 9,000 clips} \\
\midrule

target categories & 397 \\
Prompt & text and/or 1\,s audio cue \\
Mixture & 1--2 targets/mixture; active/frame $\in\{0,1,2\}$ \\
Noise & signal-to-noise ratio (SNR) $\sim \mathcal{U}([-5,5])$ dB \\
\midrule

Clip format & 5\,s, dual channel, 16\,kHz \\
Labeling &
\makecell[l]{Azimuth bins $A{=}180$ over $[0,180)$\\
Label rate 15\,Hz ($T_{\mathrm{out}}{=}75$)} \\
\bottomrule
\end{tabularx}
\end{table}

Our model is trained and evaluated on a synthesized dataset that covers a wide range of acoustic conditions. The synthesis process combines clean source signals with simulated dynamic RIRs and ambient noise.

\paragraph{Sound Sources}
To ensure diversity in our target sound events, we collected source signals from several public datasets. Specifically, speech signals were collected from the LibriSpeech corpus~\cite{panayotov2015librispeech}. Musical instrument sounds were collected from the CC-Music Pianos~\cite{zhou2023holistic} and GuitarSet datasets~\cite{xi2018guitarset}. For a broader range of general acoustic events, we utilized samples from AudioSet~\cite{gemmeke2017audio} and WavCaps~\cite{mei2024wavcaps}.

\paragraph{Noise Sources}
A collection of noise recordings was used to create realistic and challenging noisy mixtures. These were sourced from multiple datasets, including MS-SNSD~\cite{reddy2020interspeech}, WHAM!~\cite{wichern2019wham}, ESC-50~\cite{piczak2015dataset}, UrbanSound~\cite{Salamon:UrbanSound:ACMMM:14}, QUT-NOISE~\cite{dean2010qut} and Musan~\cite{snyder2015musan}.

\paragraph{Data Simulation}
We generated all training, validation, and test mixtures synthetically. The process relies on dynamic RIRs generated using the GPURIR library~\cite{diaz2021gpurir}. For each simulated scenario, we define a room with dimensions of $4 \times 4 \times 2$ meters and a reverberation time ($T_{60}$) of $0.2$ seconds. A dual-channel microphone array is positioned at coordinates $[1.9, 2.0, 1.0]$ and $[2.1, 2.0, 1.0]$, with the inter-microphone distance of 20 cm.

To simulate moving sources, we generate 5-s trajectories within a frontal azimuth range of $180^\circ$ and uniformly resample them at $15 \ \text{Hz}$, yielding $T_{\mathrm{out}} = 75$ discrete target directions per clip. Motivated by the DCASE SELD benchmarks, we use a label frame rate coarser than the acoustic feature frame rate to shorten sequences while retaining sufficient temporal detail~\cite{politis2020dataset}. At $15 \ \text{Hz}$, a source traversing $180^\circ$ in 5-s moves by at most $2.4^\circ$ between consecutive labels, which is much smaller than the $20^\circ$ angular tolerance adopted in the DCASE SELD localization metrics~\cite{politis2020dataset,politis2020overview}.

Each mixture is generated with one or two moving sound sources. Since these sources may be intermittently silent, the number of active sources in any given time frame is zero, one, or two. These moving sources are created by convolving the clean source signals with their corresponding dynamic RIRs. Finally, a randomly selected noise signal is added to the mixture at SNR uniformly sampled from the range of $[-5, 5]$ dB.
In total, we generated 288.9 hours of audio for the training set, 31.1 hours for the validation set, and 18.1 hours for the test set. 

\paragraph{Real Recordings}
\label{tau}
We use a subset of nine rooms from TAU-SRIR~\cite{tau_srir_2022}: Bomb shelter, Gym, PB132, PC226, SA203, SC203, SE203, TB103, and TC352. This subset is chosen to cover the main indoor archetypes represented in the dataset, span distinct surface materials and room volumes (rock and plastic-coated surfaces, carpet vs.\ hard floors, glass partitions), and include both trajectory types provided by the dataset (circular in Bomb shelter/Gym/PB132/PC226 and linear in SA203/SC203/SE203/TB103/TC352). Although TAU-SRIR provides discrete source positions along circular/linear trajectories, we render each 5-s evaluation segment with a single spatial room impulse response (SRIR), i.e., a measured multichannel RIR with fixed source--microphone geometry, selecting the symmetric mic pair from the tetrahedral array to obtain two channels. The goal is to ensure diverse reverberation and reflection patterns while keeping the evaluation compact and non-redundant. Clean 5-s utterances from LibriSpeech~\cite{panayotov2015librispeech} are convolved with the selected two-channel RIRs (symmetric mic pair 0 and 2) and mixed with background noise at SNR uniformly sampled from $[-5,5]\,\mathrm{dB}$; angles are expressed in the two-microphone frame by subtracting the array baseline and folding to $[0^\circ,180^\circ)$.

\vspace{-2mm}
\subsection{Implementation details}
With a Hann STFT at $16$\,kHz ($n_{\mathrm{fft}}=1024$, hop $=256$) we obtain $F=513$ and $T_{\mathrm{in}}=251$; in the audio encoder, $D_{\mathrm{emb}}=256$; in the PGSA module, $K=80$, $N_{\mathrm{dprnn}}=6$, $L_{\mathrm{chunk}}=128$, $D_h = D_c=d_k=64$, $K_{\mathrm{EIE}}=2$; in the DoA estimator, $C_{\mathrm{feat}}=2180$, $C_{\mathrm{tcn}}=256$, $D_{\mathrm{gru}}=256$ (thus $2D_{\mathrm{gru}}=512$), $T_{\mathrm{out}}=75$, $\Theta=180$, $N_{\max}=2$. For training, we use up to 200 epochs with early stopping (patience $=10$) and Adam optimization with $\text{lr}=5\times10^{-4}$, and gradient clipping with $\text{clip\_norm}=1$.
For the loss weights in the training objective, we use
$(\omega_{\mathrm{sel}},\omega_{\mathrm{DoA}},\omega_{\mathrm{card}})=(1,100,1)$
in all experiments.

\vspace{-2mm}
\subsection{Baseline Methods}
\label{sec:baseline_methods}

We categorize the baselines into four types:
(1) \textbf{Track-wise methods} (IPDNet~\cite{wang2024ipdnet}, EINV2~\cite{mu2024mff}, embed-ACCDoA~\cite{shimada2024zero}, SALSA-Lite~\cite{nguyen2022salsa_lite}, the DCASE 2025 Task 3 baseline (denoted as DCASE25 below)~\cite{shimada2025stereo}), which allocate a fixed number of output tracks for DoA/activity trajectories;
(2) \textbf{SELD+DoA methods} (SELDnet~\cite{adavanne2018sound}, SELDT~\cite{adavanne2019localization}), which output per-class predictions without explicit track control;
(3) \textbf{Pure DoA methods} (SRP-DNN~\cite{yang2022srp}, FN-SSL~\cite{wang2023fn}), which estimate DoA from spatial features without activity branches; and
(4) \textbf{Prompt-based methods}, where we compare the text-queried SEL model~\cite{zhao2024text} against our proposed SelectTSL.

For fair comparison, we adapt all baselines to our binaural two-channel setup while preserving their original prediction heads and training objectives as much as possible.
Specifically, we adapt the baselines as follows:
(i) multi-microphone or FOA inputs are reduced to the L-R pair, consistent with stereo/variable-array settings in prior work~\cite{wilkins2023two,shimada2025stereo,wang2024ipdnet};
(ii) standard FOA features are replaced by two-microphone spatial cues ($\cos(\mathrm{IPD})$, $\sin(\mathrm{IPD})$, ILD, and GCC-PHAT), following established feature substitution protocols used in DCASE baselines and SALSA-lite~\cite{politis2020dataset,zhang2021data,nguyen2022salsa_lite};
(iii) the geometry is constrained to the horizontal plane ($0^\circ$--$180^\circ$ azimuth) to remove front-back ambiguity~\cite{wilkins2023two,shimada2025stereo};
(iv) original prediction heads and losses are preserved with minimal changes, in line with community practice and variable-array SSL designs~\cite{politis2020dataset,wang2024ipdnet} and
(v) inference outputs are mapped to a common frame-wise DoA heatmap format for fair comparison~\cite{politis2022starss22,nguyen2022salsa_lite}.

\vspace{-2mm}
\subsection{Metrics}
We report static frame-level metrics (MAE, Precision, F1, Recall) and dynamic trajectory-level metrics (MOTA$^\ast$, DetA, OSPA-T).
The evaluation is based on framewise DoA matching. For each frame $t$, let $\mathcal{P}_t=\{\hat\theta_{t,k}\}$ denote the set of predicted azimuth hypotheses and $\mathcal{G}_t=\{\theta_{t,k}\}$ the annotated references.
We match $\mathcal{P}_t$ to $\mathcal{G}_t$ via a Hungarian assignment that minimizes the circular distance on $[0,180^\circ)$, i.e., $d_\circ(\hat\theta,\theta)=\min\big(|\hat\theta-\theta|,\ 180^\circ-|\hat\theta-\theta|\big)$.
A match is a TP if its error $\le \delta$; unmatched predictions/references are FP/FN.
Let $\mathrm{TP}=\sum_t \mathrm{TP}_t$, $\mathrm{FP}=\sum_t \mathrm{FP}_t$, $\mathrm{FN}=\sum_t \mathrm{FN}_t$, and $\mathrm{GT}=\sum_t|\mathcal{G}_t|$.
We report precision $P=\frac{\mathrm{TP}}{\mathrm{TP}+\mathrm{FP}}$, recall $R=\frac{\mathrm{TP}}{\mathrm{TP}+\mathrm{FN}}$ and $\mathrm{F1}=\frac{2PR}{P+R}$.

\textbf{MAE} (Mean Angular Error):
$$
\mathrm{MAE}
=\frac{1}{\sum_{t} |M_t|}
\sum_{t}\ \sum_{(\hat\theta,\theta)\in M_t} d_\circ(\hat\theta,\theta),
$$
where $M_t$ denotes the Hungarian assignment between $\mathcal{P}_t$ and $\mathcal{G}_t$ in frame $t$.

\textbf{MOTA$^\ast$} (Multiple Object Tracking Accuracy; ID-agnostic):
We follow \cite{adavanne2021differentiable} but omit the ID–switch penalty; since all DoA hypotheses extracted under the same prompt belong to a single class, we report an ID-agnostic variant.

$$
\mathrm{MOTA}^\ast=1-\frac{\sum_t(\mathrm{FN}_t+\mathrm{FP}_t)}{\sum_t|\mathcal{G}_t|}=1-\frac{\mathrm{FN}+\mathrm{FP}}{\mathrm{GT}}.
$$

\textbf{DetA} (Detection Accuracy):
$$
\mathrm{DetA}=\frac{\mathrm{TP}}{\mathrm{TP}+\mathrm{FP}+\mathrm{FN}}.
$$

\textbf{OSPA-T} (Optimal Sub-Pattern Assignment~\cite{garcia2020metric}).
Let $d_c(\hat\theta,\theta)=\min\{d_\circ(\hat\theta,\theta),c\}$ and $n_t=\max(|\mathcal{P}_t|,|\mathcal{G}_t|)$.
The frame score is
$$
\mathrm{OSPA}_t=\frac{1}{n_t}\Big(\sum_{(\hat\theta,\theta)\in M_t} d_c(\hat\theta,\theta)+c(\mathrm{FP}_t+\mathrm{FN}_t)\Big),
$$
where the cutoff $c$ in~\cite{garcia2020metric} is set to $15^\circ$,
and $\mathrm{FP}_t=|\mathcal{P}_t|-|M_t|$, $\mathrm{FN}_t=|\mathcal{G}_t|-|M_t|$.
The sequence score averages over all frames:
$$
\mathrm{OSPA\text{-}T}=\frac{1}{T}\sum_{t=1}^{T}\mathrm{OSPA}_t.
$$

\section{Experiments}

{
\setlength{\textfloatsep}{6pt}
\setlength{\floatsep}{6pt}
\setlength{\abovecaptionskip}{2pt}
\setlength{\belowcaptionskip}{0pt}

\begin{table*}[t]
\centering
\caption{
Static metrics are frame-level (peak-matching); dynamic metrics are trajectory-level.
MAE is the mean angular error on true positives. Prec. denotes precision.
MOTA$^\ast$ is ID-agnostic (no ID-switch penalty). OSPA-T is reported in degrees with $p{=}1$ and $c{=}15$.
Prec./F1/Recall/MOTA$^\ast$/DetA are reported as percentages.
\textsuperscript{\dag}\,Non–track-wise: the model cannot specify the number of tracks when producing DoA estimates.
Global best results are in bold; Category best results are underlined.
}
\label{tab:results_combined}
\small
\setlength{\tabcolsep}{3pt}
\renewcommand{\arraystretch}{0.95}
\begin{tabularx}{\textwidth}{@{} l L c c c c c c c c @{}}
\toprule
\textbf{Category} & \textbf{Model} & \textbf{Input} &
\multicolumn{4}{c}{\textbf{Static DoA (frame-level)}} &
\multicolumn{3}{c}{\textbf{Dynamic metrics (trajectory-level)}} \\
\cmidrule(lr){4-7}\cmidrule(l){8-10}
 &  &  & MAE (°)\,$\downarrow$ & Prec. (\%)\,$\uparrow$ & F1 (\%)\,$\uparrow$ & Recall (\%)\,$\uparrow$ & MOTA$^\ast$ (\%)\,$\uparrow$ & DetA (\%)\,$\uparrow$ & OSPA-T (°)\,$\downarrow$ \\
\midrule

\multirow{15}{*}{\textbf{Track-wise}}
  & \multirow{3}{*}{IPDNet}
    & Mix & 4.60 & 74.69 & 23.63 & 15.62 & 21.88 & 27.03 & 11.09 \\
\rowcolor{gray!6}
  &  & Sel-Joint & 4.17 & 78.00 & 42.69 & 29.39 & 21.10 & 27.14 & 10.23 \\
  &  & Clean & \catbest{1.40} & \catbest{82.62} & 81.07 & 90.83 & \catbest{69.33} & 72.69 & 4.55 \\
\cmidrule(lr){2-10}
  & \multirow{3}{*}{EINV2}
    & Mix & 11.37 & 47.27 & 40.36 & 31.16 & 7.91 & 25.28 & 11.26 \\
\rowcolor{gray!6}
  &  & Sel-Joint & 4.18 & 69.12 & 63.99 & 49.50 & 41.04 & 52.40 & 7.66 \\
  &  & Clean & 1.45 & 76.25 & \catbest{84.80} & \bestcat{95.52} & 65.77 & \catbest{73.62} & \catbest{4.15} \\
\cmidrule(lr){2-10}
  & \multirow{3}{*}{embed-ACCDoA}
    & Mix & 20.02 & 38.33 & 1.17 & 0.59 & -0.36 & 0.59 & 14.74 \\
\rowcolor{gray!6}
  &  & Sel-Joint & 6.28 & 59.96 & 34.20 & 23.93 & 7.95 & 20.63 & 10.86 \\
  &  & Clean & 6.51 & 60.00 & 39.83 & 29.80 & 9.93 & 24.86 & 9.88 \\
\cmidrule(lr){2-10}
  & \multirow{3}{*}{SALSA-Lite}
    & Mix & 19.66 & 36.22 & 1.36 & 0.69 & -0.53 & 0.68 & 14.72 \\
\rowcolor{gray!6}
  &  & Sel-Joint & 4.66 & 70.12 & 31.53 & 20.34 & 11.67 & 18.72 & 11.46 \\
  &  & Clean & 4.34 & 72.60 & 37.08 & 24.89 & 15.50 & 22.76 & 10.78 \\
\cmidrule(lr){2-10}
  & \multirow{3}{*}{DCASE25}
    & Mix & 17.15 & 43.34 & 0.69 & 0.35 & -0.11 & 0.35 & 14.77 \\
\rowcolor{gray!6}
  &  & Sel-Joint & 5.71 & 63.57 & 35.07 & 24.21 & 10.34 & 21.26 & 10.79 \\
  &  & Clean & 5.80 & 64.40 & 42.06 & 31.23 & 13.97 & 26.63 & 9.72 \\
\midrule

\multirow{6}{*}{\textbf{SELD+DoA}\textsuperscript{\dag}}
  & \multirow{3}{*}{SELDnet}
    & Mix & 15.83 & 42.62 & 1.51 & 0.77 & -0.27 & 0.76 & 14.71 \\
\rowcolor{gray!6}
  &  & Sel-Joint & 6.62 & 57.66 & 35.18 & 25.31 & 6.72 & 21.34 & 10.54 \\
  &  & Clean & \catbest{6.34} & 60.76 & 41.36 & 41.36 & 11.10 & 26.08 & 9.65 \\
\cmidrule(lr){2-10}
  & \multirow{3}{*}{SELDT}
    & Mix & 22.12 & 31.81 & 2.97 & 1.56 & -1.78 & 1.51 & 14.58 \\
\rowcolor{gray!6}
  &  & Sel-Joint & 6.66 & 61.53 & 48.26 & 39.71 & 14.88 & 31.81 & 8.40 \\
  &  & Clean & 6.63 & \catbest{62.96} & \catbest{53.17} & \catbest{46.01} & \catbest{18.94} & \catbest{36.21} & \catbest{7.49} \\
\midrule

\multirow{6}{*}{\textbf{Pure DoA}\textsuperscript{\dag}}
  & \multirow{3}{*}{SRP-DNN}
    & Mix & 6.24 & 59.91 & 31.34 & 21.22 & 6.98 & 18.58 & 11.21 \\
\rowcolor{gray!6}
  &  & Sel-Joint & \catbest{5.56} & 63.83 & 26.49 & 16.71 & 7.24 & 15.27 & 11.99 \\
  &  & Clean & 5.97 & 62.86 & \catbest{38.63} & \catbest{27.89} & 11.41 & 23.94 & 10.17 \\
\cmidrule(lr){2-10}
  & \multirow{3}{*}{FN-SSL}
    & Mix & 7.01 & 70.23 & 17.41 & 19.08 & 18.23 & 17.87 & 10.09 \\
\rowcolor{gray!6}
  &  & Sel-Joint & 5.83 & \catbest{82.56} & 35.12 & 22.98 & \catbest{23.10} & \catbest{25.13} & \catbest{7.82} \\
  &  & Clean & 5.92 & 79.72 & 33.82 & 24.29 & 21.13 & 23.78 & 8.84 \\
\midrule

\multirow{4}{*}{\textbf{Prompt-based}}
  & \multirow{3}{*}{SEL}
    & Mix & 5.37 & 59.82 & 30.45 & 20.42 & 6.70 & 7.96 & 13.14 \\
\rowcolor{gray!6}
  &  & Sel-Joint & 2.78 & 83.01 & 49.50 & 44.17 & 16.25 & 23.71 & 7.96 \\
  &  & Clean & 2.84 & 82.91 & 50.23 & 43.88 & 17.14 & 21.17 & 8.01 \\
\cmidrule(lr){2-10}
  & Ours & Mix & \bestcat{0.98} & \bestcat{98.28} & \bestcat{95.67} & \catbest{93.20} & \bestcat{91.57} & \bestcat{91.70} & \bestcat{2.08} \\
\bottomrule
\end{tabularx}
\end{table*}

}

\subsection{Main Results}
Table~\ref{tab:results_combined} reports the performance of all baselines
across static frame-level metrics (MAE, Prec., F1, Recall) and dynamic trajectory-level
metrics (MOTA$^\ast$, DetA, OSPA-T).
For readability, we discuss results by the four baseline families defined in Sec.~\ref{sec:baseline_methods}.
To disentangle architectural effects from input conditioning, we report results under three input conditions: Mix, Sel-Joint, and Clean.
Specifically, Mix denotes the original stereo mixture with an unknown number of active sources (0/1/2) plus noise, fed directly to the baseline without any extraction front-end.
Sel-Joint denotes end-to-end training where the baseline is preceded by the same PGSA module as ours and takes the concatenation of the PGSA-enhanced magnitude and the original Mix phase as inputs, forming a two-channel representation.
Clean is a clean condition where the input contains only the target source(s) consistent with the prompt, with no interfering sound sources and no noise.

\subsubsection{Track-wise baselines}

Within the track-wise family (IPDNet, EINV2, embed-ACCDoA, SALSA-Lite, DCASE25), only IPDNet and EINV2 remain effective in the Mix condition.
IPDNet attains an MAE of $4.60^\circ$ and an F1 of 0.24, whereas EINV2 achieves a higher F1 of 0.40 but at the cost of a much larger MAE of $11.37^\circ$.
This apparent trade-off (high F1 but large MAE) is consistent with the DP-IPD and ACCDoA formulations. When the target is active, the per-track activity outputs are frequently triggered, leading to many detections and thus a higher F1.
However, in the Mix condition the predicted DoAs are biased toward a direction between the target and interfering sources, which can still fall within the evaluation tolerance while increasing the average angular error.
The remaining track-wise baselines have almost zero F1 and negative or near-zero MOTA$^\ast$ on Mix, and only become viable once Sel-Joint or Clean inputs simplify the scene.

Under Clean inputs, IPDNet and EINV2 achieve MAE $\approx 1.4^\circ$ and MOTA$^\ast$ $\approx 0.66$--0.69, indicating that track-wise modeling can work once the prompt-based target is extracted.
However, these systems are optimized to explain all active sources rather than follow a single target: DP-IPD based IPDNet treats any source at a given direction similarly, and ACCDoA-style systems (EINV2, SALSA-Lite, embed-ACCDoA, DCASE25) only weakly encode the prompt.
In complex mixtures, tracks tend to follow non-target sources with similar geometry or break the target trajectory into short, fragmented segments.
Our model directly targets the key weakness of generic track-wise designs: weak prompt binding that leads to drift and trajectory fragmentation in complex mixtures.
We enforce target selectivity with a prompt-based PGSA front-end, whose EIEs condition the IPD Enhancer so that the refined spatial cues are biased toward the prompt-specified source.
On top of these target-dominant cues, we predict a per-azimuth DoA posteriorgram together with a cardinality head, avoiding slot-based track competition and stabilizing tracking under time-varying activity, which yields much higher MOTA$^\ast$ on Mix (Table~\ref{tab:results_combined}).

\subsubsection{SELD+DoA (non-track-wise)}

The SELD+DoA baselines (SELDnet, SELDT) predict frame-level activity and DoA per class without explicit tracks.
In the Mix condition they are highly sensitive to interference: SELDnet attains MAE $\approx 15.8^\circ$, SELDT MAE $\approx 22.1^\circ$, and both have MOTA$^\ast\!\leq 0$ with very poor detection.
With Sel-Joint and Clean inputs, their static metrics improve substantially (MAE $\approx 6.6^\circ$), consistent with prior SELD results on less cluttered scenes.

Trajectory-level metrics improve much less.
Even in Clean, SELDnet remains below MOTA$^\ast=0.12$ and SELDT only reaches MOTA$^\ast\approx 0.19$, suggesting that temporal inconsistency in the frame-level DoA predictions still leads to fragmented trajectories.
In comparison, SELDT is clearly stronger on trajectory-level metrics, confirming that its tracking-oriented objectives help stabilize trajectories.
However, both models are trained to explain all event classes rather than a single prompt-based target. When multiple sources overlap, they distribute energy across several classes and directions instead of committing to a single, stable trajectory for the target.
Consequently, even with Sel-Joint or Clean inputs their MOTA$^\ast$ remains far below both the best track-wise baselines and our SelectTSL, which explicitly enforces selectivity and temporal consistency with respect to prompt.

\subsubsection{Pure DoA (non-track-wise)}

The pure DoA baselines (SRP-DNN, FN-SSL) operate without any sound event detection (SED) head or explicit tracks, estimating DoA directly from binaural phase cues.
SRP-DNN uses a causal CRNN to estimate a frame-wise spatial spectrum on a fixed azimuth grid, followed by peak-picking to obtain the DoA.
In our binaural horizontal setup this yields very similar localization across input conditions (MAE = $6.24^\circ$ / Mix, $5.56^\circ$ / Sel-Joint, $5.97^\circ$ / Clean) and only modest changes in MOTA$^\ast$ ($\approx 0.07/0.07/0.11$).
This limited variation reflects the fixed spatial grid and the absence of explicit temporal modeling: cleaner magnitudes sharpen the peaks but only mildly affect their centers, while framewise fluctuations still lead to fragmented trajectories.

FN-SSL also estimates DP-IPD but uses full-band / narrow-band fusion to exploit cross-frequency and temporal structure.
From Mix to Sel-Joint and Clean, its MAE improves from $7.01^\circ$ to  $5.8^\circ$, and MOTA$^\ast$ increases from 0.18 to 0.23 and 0.21, yielding a more favorable static–dynamic trade-off than SRP-DNN.
Nevertheless, FN-SSL remains a purely geometric, single-stage DoA estimator that treats all spatial peaks in the mixture as equally plausible, without any mechanism to focus on the target source.
Consequently, it significantly underperforms our SelectTSL integrating the PGSA front-end. This performance gap across both static and trajectory-level metrics confirms that geometry alone is insufficient for robust, prompt-based localization and tracking.

\subsubsection{Prompt-based}

The prompt-based baseline SEL~\cite{zhao2024text} conditions localization directly on a text description of the target event.
It takes multichannel audio and a text query, fuses an audio representation with a CLAP-based text embedding, and predicts a discretized azimuth distribution per frame.
Under the Mix condition, SEL already benefits from text conditioning but remains limited by interference and ambiguous audio–text correspondence (MAE = $5.37^\circ$, MOTA$^\ast$ = 0.07).

Prepending the PGSA module (Sel-Joint) or switching to Clean inputs substantially strengthens frame-level localization (MAE $\approx 2.8^\circ$ in both cases) and improves trajectory metrics (MOTA$\approx$0.16–0.17).
However, SEL predicts a single azimuth distribution per frame for a given text embedding. When more than one source matches the query, it often focuses on the dominant target or averages multiple directions, leading to missed or unstable secondary trajectories, consistent with the original benchmark.

Our model is explicitly designed to handle up to at least two concurrent target sources per text query.
Using the raw Mix as input, it achieves an MAE of $0.98^\circ$ and a MOTA$^\ast$ of 0.92, substantially outperforming SEL even when SEL is given Sel-Joint or Clean inputs.
We attribute these gains to the combination of a prompt-based PGSA front-end and a per-azimuth DoA heatmap that represents multiple peaks per frame, rather than collapsing them into a single discrete angle.
This yields much sharper and more stable trajectories, especially in frames where two target sources overlap.

\subsection{Ablation and Design Analysis of SelectTSL}

\subsubsection{System-level architectural ablations}

We perform system-level ablations to quantify the contribution of key design choices in SelectTSL (Table~\ref{tab:ablation_results}), including:
(i) coupling between selection and DoA (conditioning on extracted target magnitude vs.\ mixture magnitude),
(ii) spatial cues to the DoA estimator (IPD+ILD, w/o ILD, w/o IPD, or none),
(iii) the number $K_{\mathrm{EIE}}$ of DPRNN blocks used to form EIEs,
(iv) the DoA module design (with/without cross-attention and FuseLayer),
(v) the cardinality head, and
(vi) the training scheme (end-to-end vs.\ two-stage).
In each row, only the factor in the ``Setting'' column is changed and all other components follow the main configuration.

Across (i)--(ii), both target-dependent conditioning and explicit multichannel cues are indispensable.
Even with EIEs disabled ($K_{\mathrm{EIE}}{=}0$), conditioning on the selected target clearly outperforms the Mix$\rightarrow$DoA variant using mixture magnitude (MAE $1.21^\circ$ vs.\ $3.32^\circ$, MOTA$^\ast$ $0.88$ vs.\ $0.49$).
Removing ILD already degrades performance, while removing IPD or all spatial cues leads to large errors (MAE $\geq 3.89^\circ$) and much lower MOTA$^\ast$, confirming the central role of phase-based spatial information in reverberant multi-source scenes.

For temporal and structural choices (iii)-(vi), using EIEs from only a few DPRNN blocks is sufficient: our default $K_{\mathrm{EIE}}{=}2$ performs best, $K_{\mathrm{EIE}}{=}1$ remains close, whereas $K_{\mathrm{EIE}}{=}3$ or 4 monotonically worsen MAE and MOTA$^\ast$, suggesting that incorporating EIEs from more blocks mainly injects noise.
Removing cross-attention in the Extraction Network hurts trajectory metrics more than replacing the Fusion Layer with simple concatenation, since cross-attention is where the prompt interacts with the mixture features. Removing it weakens target emphasis and increases interference, which hurts temporal association.
The cardinality head is crucial for stable tracking: removing it sharply reduces MOTA$^\ast$ (0.89$\rightarrow$0.58) and increases OSPA (3.1$^\circ\rightarrow$6.3$^\circ$).
Finally, the two-stage training variant lags behind end-to-end optimization (MAE $1.28^\circ$, F1 $0.90$, MOTA$^\ast$ $0.82$), indicating that joint training is needed for PGSA to produce EIEs that are aligned with the DoA objective.

\begin{table}[t]
\centering
\caption{Ablation study of SelectTSL (MAE, F1, MOTA$^\ast$, OSPA-T; $\downarrow$ lower is better, $\uparrow$ higher is better). 
F1 and MOTA$^\ast$ are reported as percentages.
The Full model row is shown as a reference and is not considered when highlighting best results.
Best results within each ablation group are highlighted in bold.}
\label{tab:ablation_results}
\scriptsize
\setlength{\tabcolsep}{1.5pt}
\renewcommand{\arraystretch}{1.05}
\begin{tabular*}{\columnwidth}{@{\extracolsep{\fill}} lrrrr@{}}
\toprule
\textbf{Setting} & \textbf{MAE (°)\,$\downarrow$} & \textbf{F1 (\%)\,$\uparrow$} & \textbf{MOTA$^\ast$ (\%)\,$\uparrow$} & \textbf{OSPA-T (°)\,$\downarrow$} \\
\midrule

\rowcolor{gray!10}
\textbf{Full model (SelectTSL)}                  & 0.98 & 95.67 & 91.57 & 2.08 \\

\cmidrule(lr){1-5}
Coupling: w/o EIEs ($K_{\mathrm{EIE}}{=}0$)       & \best{1.21} & \best{93.58} & \best{87.93} & \best{2.72} \\
\rowcolor{gray!6}
Coupling: Mix$\rightarrow$DoA                    & 3.32 & 65.92 & 49.16 & 7.82 \\

\cmidrule(lr){1-5}
Spatial: w/o ILD                                   & \best{1.79} & \best{87.18} & \best{77.27} & \best{4.14} \\
\rowcolor{gray!6}
Spatial: w/o IPD                                   & 3.89 & 60.43 & 43.30 & 8.39 \\
Spatial: w/o spatial cues                        & 4.54 & 54.52 & 37.48 & 9.20 \\

\cmidrule(lr){1-5}
\rowcolor{gray!6}
EIE blocks: 1                                   & \best{1.43} & \best{91.68} & \best{88.76} & \best{3.11} \\
EIE blocks: 3                                   & 1.95 & 84.98 & 73.88 & 4.50 \\
\rowcolor{gray!6}
EIE blocks: 4                                   & 2.25 & 81.79 & 69.20 & 4.97 \\

\cmidrule(lr){1-5}
Backbone: w/o cross-attn                           & 1.50 & 91.04 & 83.55 & 3.24 \\
\rowcolor{gray!6}
Backbone: w/o FuseLayer                            & \best{1.08} & \best{94.65} & \best{89.84} & \best{2.40} \\

\cmidrule(lr){1-5}
Cardinality: w/o head                            & 2.91 & 73.57 & 58.19 & 6.33 \\

\cmidrule(lr){1-5}
\rowcolor{gray!6}
Training: separate Sel/DoA                      & 1.28 & 90.16 & 81.64 & 3.56 \\

\bottomrule
\end{tabular*}
\end{table}

\subsubsection{Prompt-only ablations}

To disentangle the roles of text and audio cues at the fusion layer, we run prompt-only ablations where the mixture is always encoded but one conditioning branch is masked to an all-zero embedding, forcing the model to rely solely on the remaining modality (Table~\ref{tab:prompt-ablation}).

For text-only conditioning, we keep the text branch active and zero out the audio embedding.
We use Qwen2.5-7B~\cite{qwen2} to generate paraphrases for each class and group them into three bands by CLAP similarity to the canonical caption: high, medium, and low ($[0.85,1.00]$, $[0.70,0.85)$, $[0.50,0.70)$).
Performance degrades monotonically as similarity decreases: relative to the canonical ``Text-Only'' caption (MAE = $1.12^\circ$, MOTA$^\ast$ = 0.86), the highest band already worsens to MAE = $1.94^\circ$/MOTA$^\ast$ = 0.74, and the lowest band falls to MAE = $3.48^\circ$/MOTA$^\ast$ = 0.43, indicating that semantically close prompts are crucial when no audio cue is available.

For audio-only conditioning, we mask the text embedding and feed a 1\,s audio cue through the audio-prompt branch.
We compare four strategies: taking the first 1\,s from the class clip (Audio-Only / Front), sampling a random 1\,s segment (Pos-Rand), time-stretching the first 1\,s (Aug-Front), and taking the first 1\,s from another clip of the same class (XClip-Front).
Using the clip onset yields the best audio-only performance (MAE = $1.57^\circ$, MOTA$^\ast$ = 0.81); random or cross-clip cues substantially degrade it (MAE $\approx 2.30^\circ$, MOTA$^\ast$ $\approx 0.51$), with simple time-stretch in between.
This suggests that under audio-only conditioning, the model is most sensitive to temporal misalignment and identity mismatch, while moderate temporal deformation is less harmful.

\begin{table}[t!]
\centering
\caption{Joint ablation over text and audio prompts under single-modality conditioning at the fusion layer. F1 and MOTA$^\ast$ are reported on a 0--1 scale.}
\label{tab:prompt-ablation}
{\setlength{\tabcolsep}{3pt}
 \renewcommand{\arraystretch}{0.9}
 \scriptsize
\begin{tabular}{@{}llcccc@{}}
\toprule
\textbf{Modality} & \textbf{Setting / Similarity} & \textbf{MAE (°)\,$\downarrow$} & \textbf{F1\,$\uparrow$} & \textbf{MOTA$^\ast$\,$\uparrow$} & \textbf{OSPA-T (°)\,$\downarrow$} \\
\midrule
\multirow{4}{*}{Text-only}
  & Text-Only              & 1.12 & 0.93 & 0.86 & 2.95 \\
  & $[0.85, 1.00]$         & 1.94 & 0.86 & 0.74 & 4.37 \\
  & $[0.70, 0.85)$         & 2.19 & 0.80 & 0.65 & 5.82 \\
  & $[0.50, 0.70)$         & 3.48 & 0.65 & 0.43 & 8.22 \\
\midrule
\multirow{4}{*}{Audio-only}
  & Audio-Only             & 1.57 & 0.89 & 0.81 & 3.67 \\
  & Pos-Rand               & 2.30 & 0.71 & 0.52 & 7.28 \\
  & Aug-Front              & 1.92 & 0.78 & 0.62 & 5.93 \\
  & XClip-Front            & 2.29 & 0.70 & 0.51 & 7.46 \\
\bottomrule
\end{tabular}%
}
\end{table}

\subsubsection{DoA estimator ablations}

We ablate the proposed DoA estimator to assess the IPD Enhancer, the semantic–acoustic branches, and the feature refinement block (FRB).
Table~\ref{tab:DoA-estimator} summarizes MAE, frame-level F1, MOTA$^\ast$, and OSPA-T; below we focus on MAE and MOTA$^\ast$ as representative static and trajectory metrics.

For the IPD Enhancer and semantic–acoustic design (A1–A4), removing the IPD Enhancer (A1) causes a marked degradation relative to the full model (MAE = $0.98^{\circ}\rightarrow2.10^{\circ}$, MOTA$^\ast$ = $0.92\rightarrow0.69$), showing that denoising and refining raw IPD is critical for robust localization.
Using only the semantic branch (A2) or only the acoustic branch (A3) is also suboptimal: both fall behind the full system, and semantic-only outperforms acoustic-only, indicating that text-conditioned embeddings carry strong class cues but still benefit from being fused with spatial features.
Replacing enhanced IPD with direct IPD input (A4) yields the worst performance in this group (MAE = $2.71^\circ$, MOTA$^\ast$ = 0.53), further highlighting the importance of the IPD enhancement module.

For the FRB variants (B1–B7), the results show that iterative refinement is necessary but must be carefully structured.
Completely removing FRB (B1) severely hurts performance (MAE = $3.80^\circ$, MOTA$^\ast$ = 0.38), indicating that a single pass through the backbone is insufficient.
Varying FRB depth (B2–B4) suggests that the default depth $\times 2$ strikes the best balance: shallower or deeper configurations do not improve MAE or MOTA$^\ast$ and can even degrade them.
Ablating internal components (B5–B7) by removing the convolution, SE branch, or residual connection again worsens both metrics, confirming that the full FRB with all three components is most effective.

\begin{table}[t!]
\centering
\caption{Ablation on DoA estimator architecture. F1 and MOTA$^\ast$ are reported on a 0--1 scale.
}
\label{tab:DoA-estimator}

\scriptsize
\setlength{\tabcolsep}{3pt}%
\renewcommand{\arraystretch}{0.9}%

\resizebox{\columnwidth}{!}{
\begin{tabular}{@{}lcccc@{}}
\toprule
\textbf{Setting} & \textbf{MAE (°)\,$\downarrow$} & \textbf{F1\,$\uparrow$} & \textbf{MOTA$^\ast$\,$\uparrow$} & \textbf{OSPA-T (°)\,$\downarrow$} \\
\midrule
\multicolumn{5}{l}{\textit{Baseline and IPD / semantic–acoustic design (A1–A4)}} \\
Full (ours)                 & 0.98 & 0.96 & 0.92 & 2.08 \\
A1: w/o IPD Enhancer        & 2.10 & 0.83 & 0.69 & 5.02 \\
A2: semantic branch only    & 1.40 & 0.91 & 0.84 & 3.32 \\
A3: acoustic branch only    & 1.62 & 0.89 & 0.80 & 3.78 \\
A4: direct IPD input        & 2.71 & 0.73 & 0.53 & 6.50 \\
\midrule
\multicolumn{5}{l}{\textit{FRB design (B1–B7)}} \\
B1: w/o FRB                 & 3.80 & 0.64 & 0.38 & 7.90 \\
B2: FRB depth $\times 1$    & 1.52 & 0.84 & 0.79 & 3.71 \\
B3: FRB depth $\times 3$    & 1.37 & 0.83 & 0.79 & 3.81 \\
B4: FRB depth $\times 4$    & 2.12 & 0.83 & 0.69 & 4.91 \\
B5: FRB conv-only, $\times 2$ (no SE) & 2.13 & 0.83 & 0.68 & 4.95 \\
B6: FRB SE-only, $\times 2$ (no conv) & 1.96 & 0.85 & 0.72 & 4.50 \\
B7: FRB w/o residual, $\times 2$      & 1.93 & 0.85 & 0.72 & 4.62 \\
\bottomrule
\end{tabular}
}
\end{table}

\subsubsection{Cardinality head ablations}

As described in Section~\ref{sec:DoA_estimator}, we decouple DoA estimation and source-count prediction: the DoA head produces a frame-level posteriorgram over $\Theta$ azimuth bins, while a separate cardinality head predicts a distribution over $\mathcal{N}=\{0,1,2\}$.
At inference, we take $\hat{n}_t$ and select the top-$\hat{n}_t$ local maxima of $\hat{\mathbf{P}}_{\mathrm{DoA}}(t,:)$ as DoA estimates (``Ours'' in Table~\ref{tab:card-head-ablation}), so the cardinality head provides a discrete prior on the number of sources without explicitly injecting cardinality embeddings into the DoA representation.

To test whether tighter coupling helps, we compare this design with two alternatives that feed cardinality information back into the DoA head.
In the ``Card-head attention'' variant, the predicted count is mapped to an embedding that queries a multi-head attention block over DoA features, and the attention output is added via a residual connection.
In the ``Embed'' variant, the 3-way cardinality probabilities are passed through a small MLP to produce an embedding that is concatenated with the DoA head input.

Table~\ref{tab:card-head-ablation} shows that both variants are worse than the simple top-$\hat{n}$ peak selection, despite using more parameters: MAE increases from 0.98$^\circ$ to 1.52$^\circ$/1.87$^\circ$ and MOTA$^\ast$ drops from 0.92 to 0.83/0.72.
Compared with the ``no cardinality head'' configuration in Table~\ref{tab:ablation_results}, all three designs benefit from having a dedicated source-count predictor, but directly injecting cardinality embeddings into DoA features tends to distort spatial structure.
The decoupled design with top-$\hat{n}$ selection therefore offers the best overall trade-off while keeping the architecture and decoding rule simple.

\begin{table}[t!]
\centering
\caption{Ablation on how the cardinality head interacts with the DoA head. All variants share the same backbone and cardinality classifier; only the way cardinality information is used differs. F1 and MOTA$^\ast$ are reported on a 0--1 scale.}
\label{tab:card-head-ablation}

\scriptsize
\setlength{\tabcolsep}{3pt}
\renewcommand{\arraystretch}{0.9}

\resizebox{\columnwidth}{!}{%
\begin{tabular}{@{}lcccc@{}}
\toprule
\textbf{Setting} & \textbf{MAE (°)\,$\downarrow$} & \textbf{F1\,$\uparrow$} & \textbf{MOTA$^\ast$\,$\uparrow$} & \textbf{OSPA-T (°)\,$\downarrow$} \\
\midrule
Ours (top-$\hat{n}$ peaks) & 0.98 & 0.96 & 0.92 & 2.08 \\
Card-head attention        & 1.87 & 0.85 & 0.72 & 4.67 \\
Embed (concat to DoA head) & 1.52 & 0.91 & 0.83 & 3.45 \\
\bottomrule
\end{tabular}%
}
\end{table}

\subsection{Robustness and Generalization}

\subsubsection{Varying acoustic complexity}

To study robustness under different acoustic conditions, we define three difficulty levels (A/B/C) by gradually enlarging the room and widening the range of reverberation time $T_{60}$; the exact parameter ranges are given in Table~\ref{tab:dynamic}.
Level~A uses compact rooms with short $T_{60}$, Level~B moderately increases both, and Level~C spans the largest, most asymmetric rooms and the longest, most variable reverberation.

As shown in Table~\ref{tab:dynamic}, performance degrades steadily from A to C.
MAE increases from 2.36$^\circ$ to 3.03$^\circ$ and MOTA$^\ast$ drops from 0.61 to 0.46, with similar trends for the other static and trajectory-level metrics.
Larger, more asymmetric rooms introduce stronger spatial ambiguities, and longer, more variable $T_{60}$ smears binaural cues, together defining the robustness envelope of our model under increasing acoustic variability.

\begin{table*}[t!]
\centering
\caption{Performance and acoustic configuration across difficulty levels. Room dimensions $L_x,L_y,L_z$ and reverberation time $T_{60}$ are sampled from the ranges shown for each level.}
\label{tab:dynamic}
{\setlength{\tabcolsep}{5pt}
 \renewcommand{\arraystretch}{1.0}
 \footnotesize
 \resizebox{\textwidth}{!}{
\begin{tabular}{@{}lccccccccccc@{}}
\toprule
\textbf{Setting} & $\boldsymbol{L_x}$ (m) & $\boldsymbol{L_y}$ (m) & $\boldsymbol{L_z}$ (m) & $\boldsymbol{T_{60}}$ (s) & \textbf{MAE ($^\circ$)} & \textbf{Prec.} & \textbf{F1} & \textbf{Recall} & \textbf{MOTA*} & \textbf{DetA} & \textbf{OSPA-T ($^\circ$)} \\
\midrule
Main  
& $4.0$ & $4.0$ & $2.0$ & $0.20$ 
& 0.98 & 0.98 & 0.96 & 0.93 & 0.92 & 0.92 & 2.08 \\
\cmidrule(lr){1-12}
A 
& $[3.6,\,4.6]$ & $[3.6,\,4.6]$ & $[1.8,\,2.3]$ & $[0.20,\,0.35]$ 
& \textbf{2.36} & \textbf{0.90} & \textbf{0.78} & \textbf{0.69} & \textbf{0.61} & \textbf{0.64} & \textbf{5.82} \\
B 
& $[3.8,\,6.0]$ & $[3.8,\,5.2]$ & $[2.2,\,3.0]$ & $[0.25,\,0.55]$ 
& 2.65 & 0.88 & 0.72 & 0.61 & 0.52 & 0.56 & 6.82 \\
C 
& $[3.5,\,7.0]$ & $[3.5,\,6.0]$ & $[2.2,\,3.0]$ & $[0.20,\,0.65]$ 
& 3.03 & 0.83 & 0.68 & 0.58 & 0.46 & 0.52 & 7.21 \\
\bottomrule
\end{tabular}
}}
\end{table*}

\subsubsection{Varying motion dynamics}

To evaluate robustness under different motion dynamics, we define four speed buckets (A--D) that control the instantaneous angular velocity of the sources.
Bucket~A corresponds to slow motion within $\pm 5^\circ$, bucket~B to moderate motion within $\pm 15^\circ$, and buckets~C and~D to increasingly rapid motion within $\pm 30^\circ$ and $\pm 50^\circ$, respectively (Table~\ref{tab:speed}).

Table~\ref{tab:speed} shows that tracking performance degrades as motion becomes faster and more irregular.
MAE remains low for slow and moderate motion (1.20$^\circ$ in A, 1.14$^\circ$ in B) but rises to 1.55$^\circ$ and 2.32$^\circ$ in buckets~C and~D, while MOTA$^\ast$ drops from 0.96 to 0.53 with similar declines in the other metrics.
Higher angular velocities induce larger frame-to-frame direction changes, breaking the temporal smoothness the model relies on, so gradual or moderate movement is handled well whereas abrupt high-speed rotations remain challenging.

\begin{table}[t!]
\centering
\caption{Performance across different motion speed buckets.}
\label{tab:speed}
{\setlength{\tabcolsep}{3pt}
 \renewcommand{\arraystretch}{0.92}
 \scriptsize
\begin{tabular}{@{}lcccccccc@{}}
\toprule
\textbf{Bucket} & \textbf{Range ($^\circ$)} & \textbf{MAE} & \textbf{Prec.} &
\textbf{F1} & \textbf{Recall} & \textbf{MOTA*} & \textbf{DetA} & \textbf{OSPA-T ($^\circ$)} \\
\midrule
A & $\pm 5$  
& 1.20 & \textbf{0.98} & \textbf{0.98} & \textbf{0.98} & \textbf{0.96} & \textbf{0.96} & \textbf{1.63} \\
B & $\pm 15$ 
& \textbf{1.14} & 0.97 & 0.95 & 0.93 & 0.90 & 0.91 & 2.29 \\
C & $\pm 30$ 
& 1.55 & 0.95 & 0.88 & 0.82 & 0.78 & 0.78 & 3.98 \\
D & $\pm 50$ 
& 2.32 & 0.88 & 0.72 & 0.61 & 0.53 & 0.57 & 6.91 \\
\bottomrule
\end{tabular}
}
\end{table}

\subsubsection{Robustness to motion and prompt conditions}

We further test robustness when spatial dynamics are present and when text prompts may be missing (Table~\ref{tab:ablation_all_in_one_balanced_nopreamble}).
For motion, we independently control movement of the background and the target, yielding four regimes: stat--stat (static noise, static source), mov--stat (moving noise, static source), stat--mov (static noise, moving source), and mov--mov (both moving), with all other factors fixed.
For prompts, we create no-prompt clips where the caption is removed but the target remains in the mixture, and vary the no-prompt proportion $r \in \{30\%, 50\%, 70\%\}$.

In the movement group, the fully static case (stat--stat) achieves the best performance (MAE = $0.15^\circ$, MOTA$^\ast$ = 0.98), background motion alone has only a mild impact (mov--stat), while moving the target (stat--mov, mov--mov) increases MAE to about $0.97^\circ$ and reduces MOTA$^\ast$ to about 0.92.
In the no-prompt group, performance degrades as the no-prompt proportion increases: MAE rises from 1.00$^\circ$ at $r=30\%$ to 2.53$^\circ$ at $r=70\%$, while MOTA$^\ast$ drops from 0.92 to 0.60. Even so, the model remains usable when captions are frequently absent, showing that it can still localize and track targets based primarily on audio cues.

\begin{table}[t!]
\centering
\caption{Ablation and robustness under different motion and prompt conditions. F1 and MOTA$^\ast$ are reported on a 0--1 scale.}
\label{tab:ablation_all_in_one_balanced_nopreamble}

\scriptsize
\setlength{\tabcolsep}{3pt}
\renewcommand{\arraystretch}{0.96}

\resizebox{\columnwidth}{!}{%
\begin{tabular}{@{}p{1.1cm}p{1.1cm}cccc@{}}
\toprule
\textbf{Category} & \textbf{Setting} & \textbf{MAE (°)\,$\downarrow$} & \textbf{F1\,$\uparrow$} & \textbf{MOTA$^\ast$\,$\uparrow$} & \textbf{OSPA-T (°)\,$\downarrow$} \\
\midrule
\multirow{4}{*}{Movement}
 & stat--stat & \textbf{0.15} & \textbf{0.99} & \textbf{0.98} & \textbf{0.45} \\
 & mov--stat  & 0.40 & 0.98 & 0.96 & 0.70 \\
 & stat--mov  & 0.97 & 0.96 & 0.92 & 2.02 \\
 & mov--mov   & 0.98 & 0.96 & 0.92 & 2.08 \\
\midrule
\multirow{3}{*}{No-prompt}
 & r=30\%     & \textbf{1.00} & \textbf{0.96} & \textbf{0.92} & \textbf{1.20} \\
 & r=50\%     & 2.22 & 0.83 & 0.67 & 3.32 \\
 & r=70\%     & 2.53 & 0.78 & 0.60 & 4.55 \\
\bottomrule
\end{tabular}%
} 
\end{table}

\subsubsection{Real-world data}

We further assess generalization on the real-room subset of TAU-SRIR, using the same protocol as in simulation.
As summarized in Table~\ref{tab:tau-real}, the mean performance across rooms is MAE = $2.62^\circ$, MOTA$^\ast$ = 0.77, and OSPA-T = $2.85^\circ$, indicating strong transfer to measured spaces.

Room-level trends correlate with the measured room attributes and trajectory design.
PB132 is a small carpeted classroom with circular trajectories, implying shorter effective reverberation and a stable direct-to-reverberant ratio (DRR); this yields easy association and correspondingly strong scores (MAE 0.82$^\circ$, MOTA$^\ast$ 0.91, OSPA-T 1.25$^\circ$).
SA203 is a lecture hall with an inclined floor and linear trajectories at multiple ranges, which introduce stronger early/late reflections and larger DRR variation, leading to increased angular bias and track fragmentation (MAE = $6.89^\circ$, MOTA$^\ast$ = 0.64, OSPA-T = $4.26^\circ$).
SE203 (a large classroom with hard floor and linear trajectories) shows a different failure mode: framewise azimuths remain sharp (MAE = $0.79^\circ$), but repeated crossings and specular clutter cause association breaks, resulting in the highest OSPA-T ($4.74^\circ$).
Overall, the model maintains accurate DoA estimates across diverse real rooms, with performance variations that can be explained by room geometry and trajectory complexity.

\begin{table}[t!]
\centering
\caption{Performance on the TAU-SRIR real-room evaluation set.}
\label{tab:tau-real}
{\setlength{\tabcolsep}{3pt}
 \renewcommand{\arraystretch}{0.9}
 \scriptsize
 \resizebox{\linewidth}{!}{
\begin{tabular}{@{}lccccccc@{}}
\toprule
Setting & \textbf{MAE ($^\circ$)} & \textbf{Prec.} & \textbf{F1} & \textbf{Recall} & \textbf{MOTA*} & \textbf{DetA} &\textbf{ OSPA-T ($^\circ$)} \\
\midrule
01\_bomb\_shelter & 2.82 & 0.86 & 0.89 & 0.92 & 0.77 & 0.80 & 2.92 \\
02\_gym           & 2.01 & 0.87 & 0.91 & 0.94 & 0.81 & 0.83 & 2.47 \\
03\_pb132         & 0.82 & \textbf{0.93} & \textbf{0.95} & \textbf{0.97} & \textbf{0.91} & \textbf{0.91} & \textbf{1.25} \\
04\_pc226         & 1.90 & 0.89 & 0.92 & 0.95 & 0.84 & 0.85 & 2.09 \\
05\_sa203         & 6.89 & 0.81 & 0.83 & 0.85 & 0.64 & 0.70 & 4.26 \\
06\_sc203         & 2.39 & 0.89 & 0.91 & 0.94 & 0.82 & 0.84 & 2.29 \\
08\_se203         & \textbf{0.79} & 0.79 & 0.81 & 0.83 & 0.61 & 0.68 & 4.74 \\
09\_tb103         & 3.37 & 0.85 & 0.88 & 0.91 & 0.75 & 0.78 & 3.04 \\
10\_tc352         & 2.63 & 0.87 & 0.90 & 0.93 & 0.79 & 0.82 & 2.62 \\
\midrule
Mean              & 2.62 & 0.86 & 0.89 & 0.92 & 0.77 & 0.80 & 2.85 \\
\bottomrule
\end{tabular}%
}}
\end{table}

\begin{figure*}[!htbp]
    \centering
    \includegraphics[width=1.0\linewidth]{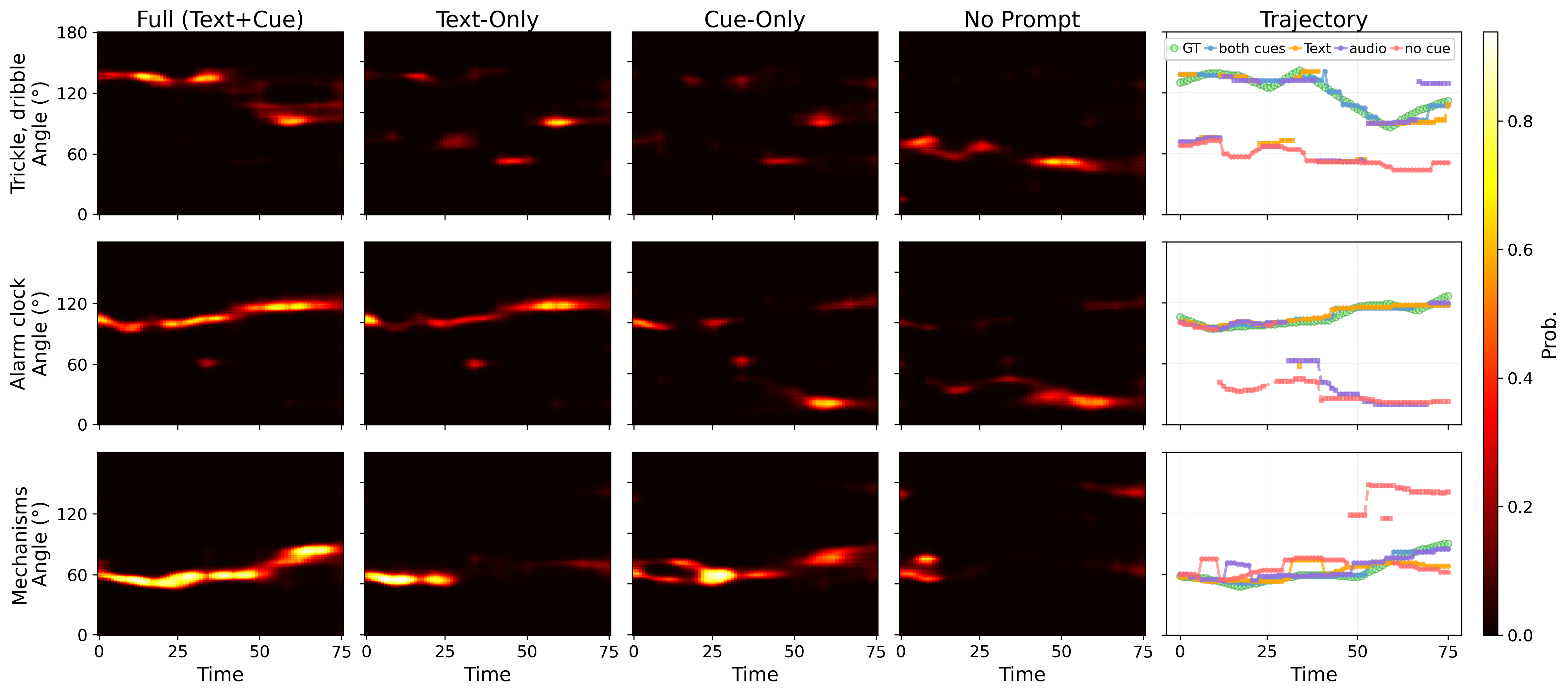}
    \caption{Heatmap visualization of frame-wise DoA posteriors from four separately trained models under different prompt configurations. For three representative target events (rows), the first four columns correspond to the Full (text+cue), Text-only, Audio-only, and No-prompt models, all evaluated on the same mixture. The rightmost column shows the estimated and GT DoA trajectories. Warmer colors indicate higher posterior probability.}
    \label{fig:heatmap}
\end{figure*}

\begin{figure*}[!htbp]
    \centering
    \includegraphics[width=1.0\linewidth]{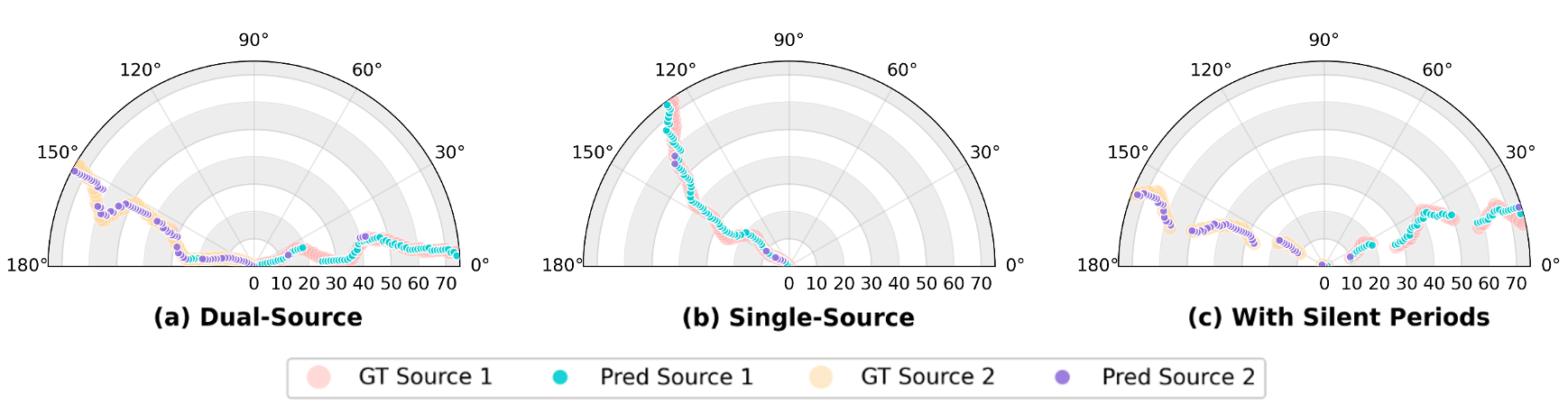}
    \caption{Polar visualization of temporal-spatial DoA trajectories. The radial axis encodes time (0--75 frames) and the angular axis encodes DoA ($0^\circ$-$180^\circ$).}
    \label{fig:DoA_traj}
\end{figure*}

\section{Visualization}
To better understand the effect of different prompt designs, we visualize the frame-wise DoA posteriors produced by four separately trained models
with different prompt inputs: (i) a Full model that uses both the audio cue and the text prompt, (ii) a Text-only model, (iii) an Audio-only model that only uses the audio cue, and (iv) a No-prompt model. For each row in Fig.~\ref{fig:heatmap}, all four models are evaluated on the same mixture recording, and the rightmost panel overlays their decoded DoA trajectories together with the GT.

In the first row, only the Full model is able to follow the challenging non-stationary trajectory, while the Text-only and Audio-only models fail to provide a reliable track and the No-prompt model produces a completely wrong trajectory. In the second row, both the Full and Text-only models closely match the ground truth, whereas the Audio-only and No-prompt models exhibit large deviations and fragmented tracks. In the third row, the Full and Audio-only models roughly capture the motion pattern, while the Text-only and No-prompt models again fail to track the source. These qualitative results demonstrate that text and audio prompts provide complementary information, and that jointly exploiting both leads to the most robust localization performance.

We also visualize estimated DoA trajectories in polar coordinates in Fig.~\ref{fig:DoA_traj}, where the radial axis encodes time (0--75 frames) and the angular axis encodes DoA ($0^\circ$-$180^\circ$).
Panel~(a) shows a dual-source recording with wide angular motion. The model follows the target across the full range.
Panel~(b) presents a single-source sequence with smooth motion, where predictions form a continuous track well aligned with the ground truth.
Panel~(c) shows a two-source sequence with silent periods that create three disjoint active segments. The model quickly re-locks onto the correct direction whenever the target becomes active again.
These examples illustrate that the model can handle dual-source, single-source, and silent-period cases while maintaining temporally consistent trajectories.



\section{Conclusion}
We presented SelectTSL, a framework for prompt-guided selective target sound localization that leverages prompt-based selectivity to learn target-aware representations for dual-channel DoA estimation.
By leveraging text and audio prompts and a target-aware multi-spatial-cue representation, our model can selectively estimate the DoA of the target source.
Experiments show consistent improvements on both static frame-level and dynamic trajectory-level metrics over competitive baselines.
Future work will extend SelectTSL toward unified multimodal prompting (e.g., incorporating visual prompts and scene context) to facilitate robust deployment in real-world environments.

\ifCLASSOPTIONcaptionsoff
  \newpage
\fi

\bibliographystyle{IEEEtran}
\bibliography{IEEEabrv,Bibliography}



\end{document}